\documentclass[a4paper,11pt]{article}
\pdfoutput=1 

\usepackage{jheppub} 
\usepackage{amsmath,amssymb,amsthm,amscd,graphicx}
\input epsf.sty

\theoremstyle{definition}

\newcommand{\CB}{{\cal B}}
\newcommand{\CC}{{\cal C}}

\newcommand{\CF}{{\cal F}}

\newcommand{\CH}{{\cal H}}
\newcommand{\CI}{{\cal I}}
\newcommand{\CJ}{{\cal J}}

\newcommand{\CO}{{\cal O}}

\def\IZ{{\mathbb Z}}
\def\IF{{\mathbb F}}
\def\IR{{\mathbb R}}
\def\IC{{\mathbb C}}
\def\IP{{\mathbb P}}

\def\IS{{\mathbb S}}
\def\IN{{\mathbb N}}

\newcommand{\tr}{{\rm Tr}}
\newcommand{\re}{{\rm e}}
\newcommand{\ri}{{\rm i}}
\newcommand{\rd}{{\rm d}}

\newcommand{\mb}{{\mathsf{b}}}
\newcommand{\map}{{\mathsf{p}}}
\newcommand{\mq}{{\mathsf{q}}}
\newcommand{\im}{{\mathsf{i}}}
\newcommand{\mg}{{\mathsf{g}}}

\def\d{\partial}

\def\tq{{\tilde{q}}}

\newcommand{\fad}{\operatorname{\Phi}_{\mathsf{b}}}

\newcommand{\fadm}{\operatorname{\Phi}_{1/\mathsf{b}}}
\newcommand{\fadi}{\fad^*}

\newcommand{\be}{\begin{equation}}
\newcommand{\ee}{\end{equation}}
\newcommand{\ba}{\begin{aligned}}
\newcommand{\ea}{\end{aligned}}
\newcommand{\ben}{\begin{eqnarray}\displaystyle}
\newcommand{\een}{\end{eqnarray}}

\newcommand{\sectiono}[1]{\section{#1}\setcounter{equation}{0}}

\newdimen\tableauside\tableauside=1.0ex
\newdimen\tableaurule\tableaurule=0.4pt
\newdimen\tableaustep
\def\phantomhrule#1{\hbox{\vbox to0pt{\hrule height\tableaurule width#1\vss}}}
\def\phantomvrule#1{\vbox{\hbox to0pt{\vrule width\tableaurule height#1\hss}}}
\def\sqr{\vbox{%
  \phantomhrule\tableaustep
  \hbox{\phantomvrule\tableaustep\kern\tableaustep\phantomvrule\tableaustep}%
  \hbox{\vbox{\phantomhrule\tableauside}\kern-\tableaurule}}}
\def\squares#1{\hbox{\count0=#1\noindent\loop\sqr
  \advance\count0 by-1 \ifnum\count0>0\repeat}}
\def\tableau#1{\vcenter{\offinterlineskip
  \tableaustep=\tableauside\advance\tableaustep by-\tableaurule
  \kern\normallineskip\hbox
    {\kern\normallineskip\vbox
      {\gettableau#1 0 }%
     \kern\normallineskip\kern\tableaurule}%
  \kern\normallineskip\kern\tableaurule}}
\def\gettableau#1{\ifnum#1=0\let\next=\null\else
\squares{#1}\let\next=\gettableau\fi\next}

\newcommand{\figref}[1]{Fig.~\protect\ref{#1}}
\title{\huge{Matrix models from operators and topological strings, 2}}

\author[a]{Rinat Kashaev,}
\author[a,b]{ Marcos Mari\~no} 
\author[b]{and Szabolcs Zakany}
\affiliation[a]{Section de Math\'ematiques, Universit\'e de Gen\`eve, Gen\`eve, CH-1211 Switzerland} 
\affiliation[b]{D\'epartement de Physique Th\'eorique,
Universit\'e de Gen\`eve, Gen\`eve, CH-1211 Switzerland}

\emailAdd{rinat.kashaev@unige.ch, marcos.marino@unige.ch, szabolcs.zakany@unige.ch}

\abstract{The quantization of mirror curves to toric Calabi--Yau threefolds leads to trace class operators, and it has been conjectured that the 
spectral properties of these operators provide a non-perturbative realization of topological string theory on these backgrounds. In this paper, we find an explicit 
form for the integral kernel of the trace class operator in the case of local $\IP^1 \times \IP^1$, in terms of Faddeev's quantum dilogarithm. The matrix model associated to this 
integral kernel is an $O(2)$ model, which generalizes the ABJ(M) matrix model. We find its exact planar limit, and we provide detailed evidence that its $1/N$ expansion 
captures the all genus topological string free energy on local $\IP^1 \times \IP^1$.}

\begin{document}
\maketitle
\flushbottom

 \sectiono{Introduction}
 
Topological strings on Calabi--Yau (CY) manifolds, just like other string theories, are only defined in perturbation theory, in terms of a genus expansion. In the 
closed string sector, the topological string free energies compute the Gromov--Witten invariants of the CY target, and for this reason topological string theory has played 
a prominent r\^ole in the interface of string theory, geometry, and mathematical physics. 

Recently, it has been conjectured in \cite{ghm} that topological strings on 
toric CY threefolds are captured, non-perturbatively, by the spectral theory of quantum-mechanical, trace class operators. These operators arise naturally in the quantization 
of their mirror curves. The conjecture of \cite{ghm} builds upon previous ideas on quantization and mirror symmetry \cite{adkmv,ns,acdkv,mirmor}, but it also 
incorporates many conceptual aspects of large $N$ dualities. In fact, many of the crucial ingredients in the proposal of \cite{ghm} were first unveiled in the 
study of the ABJM matrix model at large $N$ \cite{mp,hmo,hmo2,hmmo,kallen-m,cgm}\footnote{The approach of \cite{kallen-m} 
was first applied to topological string theory in \cite{hw}, but it requires 
corrections which are incorporated in \cite{ghm}.}. As spelled out in detail in \cite{mz}, one way 
of formulating the spectral theory/mirror symmetry correspondence of \cite{ghm} is by considering the so-called fermionic traces $Z(N, \hbar)$ of the trace class operator 
(see section \ref{gen-as} for a precise definition.) It turns out that, in the 't Hooft limit,  
\be
\label{thooftlimit}
N \rightarrow \infty, \qquad \hbar \rightarrow \infty, \qquad {N \over \hbar}=\lambda \, \, \, {\text{fixed}}, 
\ee
these traces have an asymptotic expansion of the form,  
\be
\label{free-expansion}
\log Z(N, \hbar)= \sum_{g\ge 0} \CF_g(\lambda) \hbar^{2-2g}. 
\ee
According to the conjecture of \cite{ghm}, the functions $\CF_g(\lambda)$ should be the genus $g$ free energies of the standard topological string, in the so-called 
conifold frame. The 't Hooft parameter 
$\lambda$ is a flat coordinate for the CY moduli space and is given by the vanishing period at the conifold point 
(we are assuming here that the mirror curve 
has genus one.) In this way, the weakly coupled topological 
string emerges in a limit in which the quantum-mechanical problem is strongly coupled 
(since $\hbar \rightarrow \infty$). On the other hand, the double-scaling limit 
\be
N \rightarrow \infty, \qquad \hbar \rightarrow 0, \qquad  N \hbar=\mu \, \, \, {\text{fixed}}, 
\ee
corresponds to the WKB expansion in the quantum-mechanical problem, and it is captured by the 
Nekrasov--Shatashvili (NS) limit of the refined topological string, in agreement 
with the results of \cite{mirmor,acdkv}. An obvious corollary of (\ref{free-expansion}) is that 
{\it the fermionic spectral traces $Z(N, \hbar)$ of the trace class operator provide a non-perturbative definition of 
the topological string partition function, in the spirit of large $N$ dualities.} From a more physical point of view, one can regard $Z(N, \hbar)$ as the canonical partition function of a quantum ideal 
gas of $N$ fermions, where the operator plays the r\^ole of density matrix \cite{mp}. 

As explained in \cite{mz}, if the kernel of the operator arising in the quantization of the mirror curve is known explicitly, 
then the fermionic spectral traces can be computed by a matrix model. 
Fortunately, it was shown in \cite{km} that, for some simple mirror curves (leading to so-called three-term operators), 
one can compute the corresponding kernels in closed form, in terms of Faddeev's quantum dilogarithm. This made it 
possible to verify the trace class property conjectured in \cite{ghm}. Armed with these kernels, one can compute the fermionic spectral traces $Z(N, \hbar)$, 
which are given by a generalized $O(2)$ matrix model of the type considered in \cite{kostov}. In \cite{mz} this matrix model was studied in the 
$1/N$ expansion, and it was checked in detail that, for local $\IP^2$ and a certain limit of local $\IF_2$, (\ref{free-expansion}) gives indeed the topological string free energies. 

This paper extends the results of \cite{km,mz} to an important local CY, namely local $\IP^1 \times \IP^1$. In this case, the 
mirror curve has genus one, therefore one modulus, but it also involves a mass parameter, since the geometry has two K\"ahler parameters. The quantization of 
this curve leads to a four-term operator. By using the quantum pentagon identity for Faddeev's quantum dilogarithm, 
we find an explicit expression for the integral 
kernel of the corresponding trace class operator. 
The matrix model obtained from this kernel turns out to be an $O(2)$ matrix model. We compute some spectral traces at finite $N$, 
as a function of the mass parameter, which agree with the predictions of the conjecture in \cite{ghm}, as shown in \cite{gkmr}. 
We also study the matrix model in the large $N$ limit. This can be done by doing perturbation theory in the 't Hooft coupling, as in \cite{mz}, but 
we can also use the general techniques of \cite{ek,ek2}, as developed in \cite{gm}, to solve exactly for its planar limit. 
We compare the resulting $1/N$ expansion with the topological string genus expansion, and we find a detailed agreement. 

It is known, geometrically, that the topological string on $\IP^1 \times \IP^1$ is equivalent to the topological string on local $\IF_2$ by a 
simple change of parameters \cite{gkmr} (this leads to a relation between the Gromov--Witten invariants of the two geometries, as pointed out in \cite{ikp,gkmr}). 
We show that this equivalence holds as a unitary equivalence between the corresponding 
trace class operators. This allows us to extend all of our results to local $\IF_2$ with arbitrary moduli, extending in this way the analysis presented in \cite{mz}.  

The trace class operator obtained by quantization of the mirror curve of local $\IP^1 \times \IP^1$ can be regarded 
as a generalization of the density matrix appearing in the Fermi gas formulation of the ABJ(M) 
matrix model \cite{mp,ahs,honda,honda-o}, after an analytic continuation to complex mass parameters. 
Therefore, one can rederive from our results various aspects of the ABJ(M) matrix model. 
For example, we show that the exact planar solution of the $O(2)$ matrix model reproduces the planar free energy of the ABJ(M) matrix model obtained in \cite{dmp}. 

It is natural to ask how our matrix model for local $\IP^1 \times \IP^1$ compares to a previous proposal in \cite{akmv-cs}. This proposal is based 
on a generalization of the Gopakumar--Vafa large $N$ duality \cite{gv}, in which topological string theory on local 
$\IP^1\times \IP^1$ is described by large $N$, $U(N)$ Chern--Simons theory on the lens space $L(2,1)$ \cite{akmv-cs}. When this is combined with 
the results of \cite{mmcs}, one obtains a matrix model description of topological strings on local $\IP^1 \times \IP^1$ which has been studied in 
some detail \cite{akmv-cs,hy,hoy}. There are however many important differences between these matrix models. 
First of all, the matrix model of \cite{akmv-cs} is a two-cut matrix model, while our model is a one-cut matrix model. This leads to 
important differences at the non-perturbative level, since in the model of \cite{akmv-cs} the two K\"ahler parameters of local 
$\IP^1 \times \IP^1$ are discretized (they correspond to the two filling fractions
of the two-cut matrix model), while in the matrix model described here only the ``diagonal" K\"ahler parameter is discretized. 
Another difference between these two matrix models is that the 
weak 't Hooft coupling expansion of the model in \cite{akmv-cs} corresponds 
to the so-called orbifold point in the moduli space of local $\IP^1 \times \IP^1$, while in the model 
considered here it corresponds to the conifold point. Both points lead to logarithmic periods (which 
are in fact needed to match the Gaussian behavior of the matrix models), but they are different. It would 
be interesting to understand in more detail the relationship between the two matrix models, specially at the non-perturbative 
level, but we will not pursue this problem here\footnote{We would like to thank R. Schiappa for raising this issue.}. 
Note that, if the conjecture of \cite{ghm} is true, 
the matrix model description in terms of kernels of trace class operators studied in this paper 
is likely to apply to all toric CY threefolds. In contrast, the large $N$ duality of \cite{akmv-cs} applies only to a special type of geometries, obtained as ADE quotients of 
the resolved conifold. 

This paper is organized as follows. In section 2 we elaborate on \cite{km} and obtain an explicit representation 
for the integral kernel of the trace class operator associated to local $\IP^1 \times \IP^1$. We also 
write down an $O(2)$ matrix model computing the fermionic spectral traces, and we study its $1/N$ expansion. 
We obtain perturbative results as well as a closed form 
expression for the planar free energy, which can be expanded at both weak and strong coupling. In addition, 
we show how many known results for the ABJ(M) matrix model can be recovered 
from this solution. In section 3 we compare successfully the $1/N$ expansion of the matrix model with 
the topological string free energies of local $\IP^1 \times \IP^1$, which we compute around a generic point in the 
conifold locus. We conclude in section 4 and we list some open problems for the future. In the Appendix, we 
list some properties of the quantum dilogarithm which are used in section 2.

\sectiono{Operators, kernels and matrix models}

\subsection{Integral kernel and matrix model for local $\IP^1 \times \IP^1$}
\label{gen-as}

As explained in \cite{ghm,km}, given the mirror curve to a toric CY threefold, one can 
quantize it to obtain a trace class operator $\rho$. Although this procedure can be followed for any toric geometry, the simpler case 
is that of toric (almost) del Pezzo CY threefolds, 
defined as the total space of the canonical line bundle on a toric (almost) del Pezzo surface $S$,
\be
\label{dP}
X=\CO(K_S) \rightarrow S. 
\ee
In this case, the mirror curve has genus one. The complex moduli 
of the curve involve a ``true" geometric modulus $\tilde u$ as well as a set of ``mass" parameters $m_i$, $i=1, \cdots, r$, where $r$ depends on the geometry 
under consideration \cite{hkp,hkrs}. The mirror curves can be put in the ``canonical" form
\be
\label{ex-W}
W(\re^x, \re^y)= \CO_S(x,y)+ \tilde u=0,  
\ee
where $\CO_S(x,y)$ is given by 
\be
\label{coxp}
 \CO_S (x,y)=\sum_{i=1}^{k+2} \exp\left( \nu^{(i)}_1 x+  \nu^{(i)}_2 y + f_i(m_j) \right), 
 \ee
and $f_i(m_j)$ are suitable functions of the parameters $m_j$. The vectors $\nu^{(i)}_{1,2}$ can be obtained from the toric description of the CY threefold. 
The mirror curve (\ref{ex-W}) is quantized by standard Weyl quantization. In particular, $x$, $y$ are promoted to self-adjoint Heisenberg 
operators $\mathsf{x}$, $\mathsf{y}$, satisfying the commutation relation 
\be
[\mathsf{x}, \mathsf{y}]=\im\hbar, 
\ee
and ordering ambiguities are resolved by Weyl's prescription. In this way, $\CO_S(x,y)$ becomes an operator, which 
will be denoted by $\mathsf{O}_S$. As conjectured in \cite{ghm} and proved in \cite{km} in many cases, the inverse operator
\be
\rho_S=\mathsf{O}^{-1}_S
\ee
is of trace class. 

In this paper we will focus on the important local del Pezzo CY threefold in which $S=\IP^1 \times \IP^1=\IF_0$, and usually called local $\IP^1 \times \IP^1$ or local 
$\IF_0$. Topological string theory on 
this background is known to have various applications: it engineers geometrically $SU(2)$ Seiberg--Witten theory \cite{kkv}, it is dual to 
Chern--Simons theory on the lens space $L(2,1)$ \cite{akmv-cs}, and it is closely 
related to the partition function of ABJ(M) theory on the three-sphere \cite{mp,dmp}. In this case, the function $\CO_S(x,y)$ is given by, 
\be
\label{lf0}
\CO_{\IF_0} \left(x, y \right)= \re^{ x}+ m_{\IF_0} \re^{- x} + \re^{y}  +\re^{-y}, 
\ee
and depends on a mass parameter that we denote by $m_{\IF_0}$. In principle, we will take $m_{\IF_0}$ to be real and positive, but as we will see it is possible 
to extend some of the results to complex values of $m_{\IF_0}$.

We would like to find an explicit expression for the kernel of the operator $\rho_{\IF_0}$. As for the three-term operators analyzed in \cite{km}, 
this kernel will involve in a crucial way Faddeev's quantum dilogarithm $\fad(x)$ \cite{faddeev,fk,fkv}, see the Appendix for its definition and some of 
its basic properties. In addition, the function $\fad(x)$ has the following features. If $\mathsf{p}$ and $\mathsf{q}$ are self-adjoint Heisenberg 
operators satisfying, 
 \begin{equation}
[\mathsf{p},\mathsf{q}]=(2\pi\im)^{-1}, 
\end{equation}
the quantum dilogarithm satisfies \cite{km}
\begin{equation}
\ba
\label{qd-props}
\fad(\mathsf{p})\re^{2\pi\mathsf{b}\mathsf{q}}\fadi(\mathsf{p})&=\re^{2\pi\mathsf{b}\mathsf{q}}+\re^{2\pi\mathsf{b}(\mathsf{p}+\mathsf{q})}, \\
\fadi(\mathsf{q})\fad(\mathsf{p})\re^{2\pi\mathsf{b}\mathsf{q}}\fadi(\mathsf{p})\fad(\mathsf{q})&=\re^{2\pi\mathsf{b}\mathsf{q}} +\re^{2\pi\mathsf{b}(\mathsf{p}+\mathsf{q})} 
+\re^{2\pi\mathsf{b}(\mathsf{p}+2\mathsf{q})}. 
\ea
\ee
One also has the important {\it quantum pentagon identity} \cite{faddeev-penta}, 
\be
\label{penta}
\fad(\mathsf{p})\fad(\mathsf{q})=\fad(\mathsf{q})\fad(\mathsf{p}+\mathsf{q})\fad(\mathsf{p}). 
\ee
The quantization of (\ref{lf0}) leads to the operator, 
 \begin{equation}\label{f0}
\mathsf{O}_{\mathbb{F}_0}=\re^{\mathsf{x}} +m_{\IF_0} \re^{-\mathsf{x}} +\re^{\mathsf{y}} +\re^{-\mathsf{y}}.
\ee
Let us set
\begin{equation}
\label{hb-rel}
\hbar=\pi\mathsf{b}^2  
\end{equation}
and
\be
\mathsf{x}=\pi\mathsf{b}(\mathsf{p}+2\mathsf{q}),\quad \mathsf{y}=\pi\mathsf{b}\mathsf{p}. 
\ee
By using (\ref{qd-props}), we find
 \begin{multline}
\re^{\mathsf{x}/2} \mathsf{O}_{\mathbb{F}_0}\re^{\mathsf{x}/2}-m_{\IF_0} =\re^{2\mathsf{x}} +\re^{\mathsf{x}+\mathsf{y}} 
+\re^{\mathsf{x}-\mathsf{y}}=\re^{2\pi\mathsf{b}(\mathsf{p}+2\mathsf{q})} +\re^{2\pi\mathsf{b}(\mathsf{p}+\mathsf{q})} 
+\re^{2\pi\mathsf{b}\mathsf{q}}\\
=\fadi(\mathsf{q})\fad(\mathsf{p})\re^{2\pi\mathsf{b}\mathsf{q}}\fadi(\mathsf{p})\fad(\mathsf{q}). 
\end{multline}
Therefore, 
\begin{multline}
\fadi(\mathsf{p})\fad(\mathsf{q})\re^{\mathsf{x}/2} \mathsf{O}_{\mathbb{F}_0}\re^{\mathsf{x}/2}\fadi(\mathsf{q})\fad(\mathsf{p})
=m_{\IF_0}+\re^{2\pi\mathsf{b}\mathsf{q}}=m_{\IF_0}\left(1+\re^{2\pi\mathsf{b}(\mathsf{q}-\mathsf{b}\xi /\pi)}\right)\\
=m_{\IF_0}\frac{\fad(\mathsf{q}-\mathsf{b}\xi/\pi-\im\mathsf{b}/2)}{\fad(\mathsf{q}-\mathsf{b}\xi/\pi+\im\mathsf{b}/2)},
\end{multline}
where the parameter $\xi$ is related to $m_{\IF_0}$ through the equation
\begin{equation}
\label{mmu}
m_{\IF_0}=\re^{2 \mathsf{b}^2 \xi}.
\end{equation}
Let us now define the operator
\begin{equation}\label{bop}
\mathsf{B}\equiv \fadi(\mathsf{q}-\mathsf{b}\xi/\pi-\im\mathsf{b}/2)\fadi(\mathsf{p})\fad(\mathsf{q})\re^{\pi\mathsf{b}\mathsf{p}/2}\re^{\pi\mathsf{b}\mathsf{q}}. 
\end{equation}
We obtain the following formula \cite{km}
\begin{equation}
\mathsf{O}^{-1}_{\mathbb{F}_0}=m_{\IF_0}^{-1} \mathsf{B}^*\mathsf{B}.
\end{equation}
On the other hand, we can use the quantum pentagon relation (\ref{penta}) to write the operator $\mathsf{B}$ as 
\begin{multline}
\re^{-(\pi\mathsf{b}/2)^2/(4\pi\im)}  \fad(\mathsf{p})\mathsf{B} 
=\fadi(\mathsf{p}+\mathsf{q}-\mathsf{b}\xi/\pi-\im\mathsf{b}/2)\fadi(\mathsf{q}-\mathsf{b}\xi/\pi-\im\mathsf{b}/2)\fad(\mathsf{q})\re^{\pi\mathsf{b}(\mathsf{p}+\mathsf{q})/2} \re^{\pi\mathsf{b}\mathsf{q}/2}\\
=\fadi(\mathsf{p}+\mathsf{q}-\mathsf{b}\xi/\pi-\im\mathsf{b}/2)\re^{\pi\mathsf{b}(\mathsf{p}+\mathsf{q})/2} \fadi(\mathsf{q}-\mathsf{b}\xi/\pi-\im\mathsf{b}/4)\fad(\mathsf{q}+\im\mathsf{b}/4)\re^{\pi\mathsf{b}\mathsf{q}/2}.
\end{multline}
If we introduce new momentum and position operators by
\begin{equation}
\mathsf{p}'\equiv \mathsf{p}+\mathsf{q}-\mathsf{b}\xi/\pi, \qquad \mathsf{q}'\equiv \mathsf{q}-\mathsf{b}\xi/2\pi, 
\end{equation}
we find
\be
\label{rho-ex}
\rho_{\mathbb{F}_0} =\re^{- \mathsf{b}^2 \xi /2} f(\mathsf{q}') {1\over  2\cosh(\pi\mathsf{b}\mathsf{p}')} f^*(\mathsf{q}'), \ee
where
\be
f(\mathsf{q}) =\re^{\pi \mathsf{b} \mathsf{q}/2}{\fad(\mathsf{q}-\mathsf{b}\xi/2\pi+ \ri \mathsf{b}/4)\over \fad(\mathsf{q}+\mathsf{b}\xi/2\pi- \ri \mathsf{b}/4)}.
\ee
In the position representation for the operators $\mathsf{p}'$ and $\mathsf{q}'$,
we obtain the integral kernel,
\begin{equation}
\label{f0ker}
\rho_{\IF_0}(x_1, x_2)=\langle x_1| \mathsf{O}^{-1}_{\mathbb{F}_0}|x_2\rangle=
\re^{- \mathsf{b}^2 \xi /2}   {f(x_1) f^*(x_2) \over  2\mathsf{b} \cosh\left(\pi {x_1- x_2 \over \mathsf{b}}  \right) }. 
\end{equation} 
As shown already in \cite{km}, this is a positive-definite, trace class operator on $L^2(\IR)$. Note that $\rho_{\IF_0} (x_1, x_2)$ is related by 
a unitary transformation to the symmetric, real kernel 
\be
\re^{- \mathsf{b}^2 \xi /2}   {|f(x_1)| |f(x_2)| \over  2\mathsf{b} \cosh\left(\pi {x_1- x_2 \over \mathsf{b}}  \right) }, 
\ee
which is of the type considered in \cite{zamo,tw}. In particular, as shown in these references, its diagonal resolvent can be obtained from a TBA-like system of non-linear 
integral equations. 

The spectral information of a trace class operator $\rho$ depending on a parameter $\hbar$ and 
acting on a Hilbert space $\CH$ can be encoded in different ways. The {\it spectral traces} of $\rho$ are defined by 
\be
\label{s-traces}
Z_\ell = \tr_\CH\, \rho^\ell, \qquad \ell=1, 2, \cdots
\ee
The {\it fermionic spectral traces} are given by  
\be
Z(N, \hbar) = \tr_{\Lambda^N\left(\CH \right)} \,  \left(\Lambda^N(\rho)\right),  \qquad N=1,2, \cdots, 
\ee
where the operator $\Lambda^N(\rho)$ is defined by $\rho^{\otimes N}$ acting on $\Lambda^N\left(\CH \right)$. The generating function of the fermionic spectral traces is 
the {\it Fredholm} or {\it spectral determinant} of $\rho$: 
\be
\label{f-det}
\Xi(\kappa, \hbar)= {\rm det}(1+ \kappa  \rho)=1+\sum_{N=1}^\infty Z(N, \hbar) \kappa^N, 
\ee
and is an entire function of $\kappa$ due to the trace class property of $\rho$ \cite{simon}. 
A well-known theorem of Fredholm (see chapter 3 of \cite{simon} for a proof) states that $Z(N, \hbar)$ has the matrix-model-like representation 
\be
\label{znmm}
Z(N, \hbar)= {1 \over N!}  \int  \rd ^N x \, {\rm det}\left( \rho(x_i, x_j) \right). 
\ee
In this equation, $\rho(x_1, x_2)$ is the integral kernel of the operator $\rho$, 
\be
\rho(x_1, x_2) = \langle x_1 |\rho |x_2 \rangle.  
\ee
The spectral traces (\ref{s-traces}) and the fermionic spectral traces are closely related, since one has that 
\be
\CJ(\kappa)= \log\, \Xi(\kappa, \hbar)=-\sum_{\ell=1}^\infty {Z_\ell \over \ell} \left(-\kappa\right)^\ell. 
\ee
The above quantities can be interpreted, more physically, in terms of an ideal Fermi gas of $N$ particles, as in \cite{mp}. In this setting, $\rho$ is the canonical density matrix, 
$Z(N, \hbar)$ is the canonical partition function of the gas, $\Xi(\kappa, \hbar)$ is the grand canonical partition function, and $\CJ(\kappa)$ is the grand potential. 

Since we have an explicit formula for the integral kernel of $\rho_{\IF_0}$, we can write down an explicit expression for the integral (\ref{znmm}). 
By using Cauchy's identity, as in \cite{kwy2,mp,mz}, 
 \be
 \label{cauchy}
 \ba
  {\prod_{i<j}  \left[ 2 \sinh \left( {\mu_i -\mu_j \over 2}  \right)\right]
\left[ 2 \sinh \left( {\nu_i -\nu_j   \over 2} \right) \right] \over \prod_{i,j} 2 \cosh \left( {\mu_i -\nu_j \over 2} \right)}  
 & ={\rm det}_{ij} \, {1\over 2 \cosh\left( {\mu_i - \nu_j \over 2} \right)}\\
 &=\sum_{\sigma \in S_N} (-1)^{\epsilon(\sigma)} \prod_i {1\over 2 \cosh\left( {\mu_i - \nu_{\sigma(i)} \over 2} \right)}, 
 \ea
  \ee
  we obtain the following matrix model representation for the fermionic traces of $\rho_{\IF_0}$, 
  \be
  \label{zf0}
  Z_{\IF_0}(N, \hbar)=
 {\re^{- \mathsf{b}^2 \xi N/2} \over N!} \int \frac{\rd^N u}{(2\pi)^N} \, \prod_{i=1}^N \left |f \left( {\mathsf{b} u_i \over 2\pi} \right) \right|^2 {\prod_{i<j} 4 \sinh^2 \left ( \frac{u_i-u_j}{2} \right ) \over \prod_{i,j} 2 \cosh \left ( \frac{u_i-u_j}{2} \right)},
  \ee
 where the variables $u_i$ are related to the original variables $x_i$ by 
\be
\label{up}
u_i= {2 \pi \over \mb}x_i.
\ee

\subsection{Relation to local $\IF_2$ and spectral traces}

It is known that topological string theory on the local $\IF_2$ geometry is closely related to topological string theory on local $\IF_0$ \cite{gkmr}. It turns out that this equivalence also holds at the level 
of the corresponding quantum operators. To see this, let us first redefine the operators appearing in (\ref{f0}) as, 
\begin{equation}\label{para}
\hbar=2\pi\mathsf{b}^2,\quad \mathsf{x}=2\pi\mathsf{b}\mathsf{q},\quad  \mathsf{y}=2\pi\mathsf{b}\mathsf{p}. 
\end{equation} 
We then have, 
\begin{multline}\label{exey}
\re^{\mathsf{x}}+ \re^{\mathsf{y}}= \re^{\mathsf{x}/2}\left(1+ \re^{\mathsf{y}-\mathsf{x}}\right)\re^{\mathsf{x}/2}=
\re^{\pi\mathsf{b}\mathsf{q}}\frac{\fad\left(\mathsf{p}-\mathsf{q}-\im\mathsf{b}/2\right)}{\fad\left(\mathsf{p}-\mathsf{q}+\im\mathsf{b}/2\right)}\re^{\pi\mathsf{b}\mathsf{q}}\\
=\fad\left(\mathsf{p}-\mathsf{q}\right)\re^{2\pi\mathsf{b}\mathsf{q}}\fad\left(\mathsf{p}-\mathsf{q}\right)^{-1}. 
\end{multline}
Therefore, 
\begin{multline}
{1\over \fad(\mathsf{p}-\mathsf{q})}\mathsf{O}_{\mathbb{F}_0} \fad(\mathsf{p}-\mathsf{q})
-\re^{2\pi\mathsf{b}\mathsf{q}}=
{1\over \fad(\mathsf{p}-\mathsf{q})}\left(m_{\IF_0}\re^{-2\pi\mathsf{b}\mathsf{q}}+\re^{-2\pi\mathsf{b}\mathsf{p}}\right)\fad(\mathsf{p}-\mathsf{q})\\
=m_{\IF_0} \re^{-\pi\mathsf{b}\mathsf{q}}{\fad(\mathsf{p}-\mathsf{q}-\im\mathsf{b}/2)\over \fad(\mathsf{p}-\mathsf{q}+\im\mathsf{b}/2)}\re^{-\pi\mathsf{b}\mathsf{q}}+
\re^{-\pi\mathsf{b}\mathsf{p}}{\fad(\mathsf{p}-\mathsf{q}-\im\mathsf{b}/2)\over \fad(\mathsf{p}-\mathsf{q}+\im\mathsf{b}/2)}\re^{-\pi\mathsf{b}\mathsf{p}}\\
=m_{\IF_0}\left( \re^{-2\pi\mathsf{b}\mathsf{q}}+\re^{2\pi\mathsf{b}(\mathsf{p}-2\mathsf{q})}\right)+
\re^{-2\pi\mathsf{b}\mathsf{p}}+\re^{-2\pi\mathsf{b}\mathsf{q}}
=(1+m_{\IF_0})\re^{-2\pi\mathsf{b}\mathsf{q}}+m_{\IF_0}\re^{2\pi\mathsf{b}(\mathsf{p}-2\mathsf{q})}+
\re^{-2\pi\mathsf{b}\mathsf{p}},
\end{multline}
or in terms of original variables
\begin{equation}\label{f0conj}
{1\over \fad(\mathsf{p}-\mathsf{q})}\mathsf{O}_{\mathbb{F}_0} \fad(\mathsf{p}-\mathsf{q})=
\re^{\mathsf{x}}+(1+m_{\IF_0})\re^{-\mathsf{x}}+m_{\IF_0}\re^{\mathsf{y}-2\mathsf{x}}+
\re^{-\mathsf{y}}.
\end{equation}
By defining new variables
\begin{equation}
\mathsf{x}'=\mathsf{x}+\nu,\quad \mathsf{y}'=\mathsf{y}-2\mathsf{x}-3\nu,\quad \nu=-{1\over4}\log(m_{\IF_0}),
\end{equation}
we rewrite \eqref{f0conj} as follows
 \begin{equation}
{1\over \fad(\mathsf{p}-\mathsf{q})}m_{\IF_0}^{-1/4}\mathsf{O}_{\mathbb{F}_0}\fad(\mathsf{p}-\mathsf{q})=
\re^{\mathsf{x}'} +(m_{\IF_0}^{1/2}+m_{\IF_0}^{-1/2})\re^{-\mathsf{x}'} +\re^{\mathsf{y}'} +\re^{-2\mathsf{x}'-\mathsf{y}'}. 
\end{equation}
We conclude that the operator $m_{\IF_0}^{-1/4}\mathsf{O}_{\mathbb{F}_0}$ is unitarily equivalent to the operator
 \begin{equation}\label{f2}
\mathsf{O}_{\mathbb{F}_2}=\re^{\mathsf{x}} +m_{\IF_2} \re^{-\mathsf{x}} +\re^{\mathsf{y}} +\re^{-2\mathsf{x}-\mathsf{y}}, 
\end{equation}
corresponding to the local $\IF_2$ geometry \cite{ghm,km}, after the substitution 
\be
\label{mf2}
m_{\IF_2}=m_{\IF_0}^{1/2}+m_{\IF_0}^{-1/2}. 
\ee
In the CY geometries, the rescaling by $m_{\IF_0}^{-1/4}$ leads, in view of (\ref{ex-W}), to the following relation between the moduli, 
\be
\label{u-moduli}
\tilde u_{\IF_2}=m_{\IF_0}^{-1/4} \tilde u_{\IF_0}. 
\ee
The relationships (\ref{mf2}), (\ref{u-moduli}) agree precisely with those found by a direct analysis of the 
topological string in these geometries \cite{gkmr}. This means in particular that any test of the conjecture of 
\cite{ghm} for local $\IF_0$ leads automatically to a corresponding test for local $\IF_2$. The unitary equivalence of the two operators also leads to the following equality of spectral traces, 
\be
\label{rel}
\tr \rho^\ell_{\IF_2}\left(m_{\IF_2}\right)=m_{\IF_0}^{\ell/4} \tr \rho^\ell_{\IF_0}(m_{\IF_0}), 
\ee
after the substitution (\ref{mf2}). 

Using the expression for the integral kernel in (\ref{f0ker}), as well as (\ref{rel}), we can in principle compute explicitly the first spectral traces. According to \cite{ghm}, we should 
expect simplifications in the so-called maximally supersymmetric case $\hbar=2 \pi$, which corresponds to 
\be
\mb={\sqrt{2}}
\ee
in (\ref{hb-rel}). For this value of $\mb$, we can use the functional equation (\ref{eq.tbshift}) satisfied by the quantum dilogarithm to obtain the following expression 
in terms of elementary functions, 
\be
\left|f(x)\right|^2 = {1\over 4 \cosh \left( {\pi \sqrt{2} \left(x-\mathsf{b}\xi/2\pi \right) \over 2} \right)\cosh \left( {\pi \sqrt{2} \left(x+\mathsf{b}\xi/2\pi  \right) \over 2} \right)}. 
\ee
After an appropriate change of variables, we obtain, 
\be
\tr  \rho_{\IF_0}={1\over 8 \pi} m_{\IF_0}^{-1/4} \int_{-\infty}^\infty {\rd u \over \cosh(u) \cosh(u-  {\sqrt{2}} \mathsf{b} \xi/2)}= {1\over 8 \pi} {\log (m_{\IF_0}) \over m_{\IF_0}^{1/2}-1}. 
\ee
The second trace is a little bit more complicated. We find
\be
\ba
\tr  \rho^2_{\IF_0}&={1\over 64 \pi^2} m_{\IF_0}^{-1/2} \int_{-\infty}^\infty  \int_{-\infty}^\infty  {\rd u \rd v\over \cosh(u) \cosh(u- {\sqrt{2}} \mathsf{b}\xi/2\pi) \cosh(v) \cosh(v-{\sqrt{2}} \mathsf{b}\xi/2\pi) \cosh(u-v)^2}\\
&={m_{\IF_0}^{-1/2}\over 16 \pi^2} \left[ \left({ \log(m_{\IF_0}) \over m_{\IF_0}^{1/2}-m_{\IF_0}^{-1/2}} +1\right)^2-1 -{\pi^2 \over \left(m_{\IF_0}^{1/4}+m_{\IF_0}^{-1/4} \right)^2} \right]. 
\ea
\ee
When $m_{\IF_0}=1$, these expressions give, 
\be
\tr  \rho_{\IF_0}(m_{\IF_0}=1)={1\over 4 \pi}, \qquad \tr \rho^2_{\IF_0}(m_{\IF_0}=1)={12-\pi^2 \over 64 \pi^2}. 
\ee
in accord with the values predicted in \cite{ghm} from the spectral theory/mirror symmetry correspondence. 

We can now use (\ref{rel}) to obtain the values of the same traces for local $\IF_2$. 
We find, 
\be
\label{trf2}
\ba
\tr \rho_{\IF_2}&={1\over 4 \pi} {\cosh^{-1}(m_{\IF_2}/2) \over {\sqrt{m_{\IF_2}-2}}}, \\
\tr \rho^2_{\IF_2}&={1\over 16 \pi^2} \left[ \left(2  {\cosh^{-1}(m_{\IF_2}/2)\over {\sqrt{ m_{\IF_2}^2-4}}} +1\right)^2 -1-{ \pi^2 \over m_{\IF_2}+2} \right]. 
\ea
\ee
We obtain, in particular
\be
\ba
\tr  \rho_{\IF_2}(m_{\IF_2}=0)&={1\over 8 {\sqrt{2}}}, \\
\tr \rho^2_{\IF_2}(m_{\IF_2}=0)&={1\over 64}\left( {4 \over \pi}-1 \right), \\
\ea
\ee
which were already obtained in \cite{km}, and
\be
\ba
\tr  \rho_{\IF_2}(m_{\IF_2}=1)&={1\over 12}, \\
\tr  \rho^2_{\IF_2}(m_{\IF_2}=1)&=\frac{1}{432} \left(\frac{12 \sqrt{3}}{\pi }-5\right).\\
\ea
\ee
It can be verified \cite{gkmr} that these values agree with the predictions of the conjecture in \cite{ghm}.

\subsection{Perturbative expansion}
We are now interested in studying the matrix integral (\ref{zf0}) in the 't Hooft limit (\ref{thooftlimit}). 
As in \cite{mz}, we should first analyze the integrand of (\ref{zf0}) when 
$\hbar$ (or equivalently $\mb$) is large. At the same time, we have to decide what is the appropriate scaling of the 
parameter $m_{\IF_0}$ appearing in the operator, as $\hbar$ becomes 
large. As it was explained in \cite{mz}, in order to recover the topological string for arbitrary mass parameter, we have to scale 
\be
\label{mscaling}
\log m_{\IF_0} \sim \hbar, \qquad \hbar \rightarrow \infty. 
\ee
We recall the variable $\xi$ is defined as
\be
\label{xidef}
\xi= { \pi \over 2 \hbar} \log m_{\IF_0}.
\ee
This is the mass variable that will be kept fixed in the 't Hooft limit. If we introduce the parameter
\be
\mg= {1\over \hbar}, 
\ee
we can write the matrix integral (\ref{zf0}) in the form 
\be
\label{zf0-bis}
Z (N,\hbar)=\frac{\re^{-\xi \lambda/(2 \pi \mg^2)}}{N!} 
 \int_{\IR^N}  { \rd^N u \over (2 \pi)^N}  \prod_{i=1}^N \re^{-{1\over \mg} V(u_i, \mg)}  \frac{\prod_{i<j} 4 \sinh \left( {u_i-u_j \over 2} \right)^2}{\prod_{i,j} 2 \cosh \left( {u_i -u_j \over 2}   \right)}, 
\ee
where
\be
\label{pot}
V(u, \mg)=-\mg \log\left |f\left (\frac{\mb u}{2\pi} \right) \right|^2.
\ee
As in \cite{mz}, we can now use the self-duality of Faddeev's quantum dilogarithm, 
\be
\fad(x)=\fadm(x), 
\ee
as well as (\ref{as-fd}), to obtain the following asymptotic expansion for small $\mg$, 
\be
V(u, \mg) \sim -\frac{u}{2\pi}-\frac{1}{\pi^2} \sum_{k \geq 0}(-4 \pi^4 \mg^2)^k \frac{B_{2k}(1/2)}{(2k)!} {\rm Im} \left[ {\rm Li}_{2-2k}(-\ri \re^{u+\xi})+{\rm Li}_{2-2k}(-\ri \re^{u-\xi}) \right].
\ee
If we write this expansion as
\be
\label{vug}
V(u, \mg)= \sum_{\ell \ge 0} \mg^{2\ell} V^{(\ell)} (u), 
\ee
we find that the leading contribution as $\mg\rightarrow 0$ is given by the``classical" potential,  
\be
\label{vo}
V^{(0)}(u)=-\frac{u}{2 \pi}-\frac{1}{\pi^2} \left( {\rm Im} \, {\rm Li}_2(-\ri \, \re^{u+\xi}) +{\rm Im} \, {\rm Li}_2(-\ri \, \re^{u-\xi})  \right). 
\ee
The matrix integral (\ref{zf0-bis}) is an $O(2)$ matrix model \cite{kostovon}, in which the inverse Planck constant $\mg$ plays the r\^ole of the string coupling constant, 
and the potential itself depends on $\mg$. In order to obtain the 't Hooft expansion of the free energy, we can use the 
asymptotic expansion of the potential (\ref{vug}). In particular, since this expansion only involves even powers of $\hbar$, 
we conclude that the matrix integral (\ref{zf0-bis}) admits a standard 't Hooft expansion, of the form 
\be
\label{thooft-f0}
F(N, \hbar)=\log Z(N, \hbar) =\sum_{g\ge 0} \hbar^{2-2g} \CF_g (\lambda, \xi), 
\ee
where $\lambda$ is the 't Hooft parameter introduced in (\ref{thooftlimit}), and $\xi$ was introduced in (\ref{xidef}). 
Note that, in the planar limit, only the classical part of the potential (\ref{vo}) contributes. By using the asymptotics of the dilogarithm, 
one finds that the classical potential behaves as
\be
\label{confinement}
V^{(0)}(u)\approx {|u| \over 2 \pi}, \qquad  |u|\rightarrow \infty, 
\ee
i.e. it is a linearly confining potential at infinity, similar to the potentials appearing in 
matrix models for Chern--Simons--matter theories \cite{mp,gm} and in other matrix integrals associated to quantized mirror curves \cite{mz}. The potential (\ref{vo}), for two values of 
$\xi$, as well as its asymptotic form (\ref{confinement}), are shown in \figref{pots}. 

\begin{figure}[h]
\center
\includegraphics[scale=0.6]{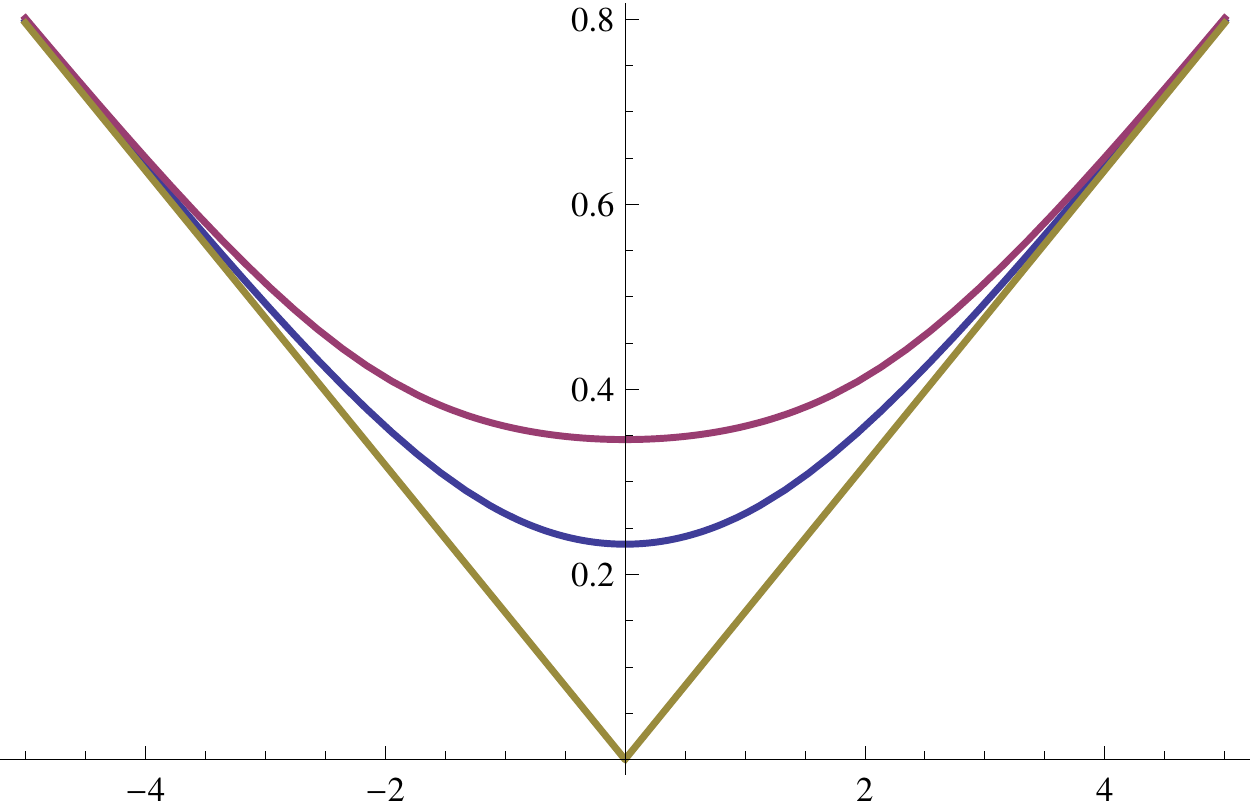}
\caption{The classical potential (\ref{vo}) as a function of $u$, for $\xi=1$ (lower line) and $\xi=2$ (upper line), together with their asymptotic form 
(\ref{confinement}) when $u$ is large.
}
\label{pots}
\end{figure}

We would like to compute the genus $g$ free energies $\CF_g (\lambda, \xi)$ appearing in the expansion (\ref{thooft-f0}). 
We will first obtain approximate expressions for the 
very first free energies, as expansions around $\lambda=0$, 
by doing perturbation theory in $\mg$, as in \cite{mz}. To do this, we regard (\ref{zf0-bis}) as a Gaussian Hermitian matrix model, perturbed by single and double trace operators. 
The computation is straightforward (see for example \cite{akmv-cs} for a similar example). For the genus zero free energy, we find the following structure, 
\be
\label{fg-stru}
\ba
	\CF_0(\lambda, \xi)&= \frac{\lambda^2}{2} \left(\log \left(\frac{\pi^2 \lambda \cosh \xi}{4} \right)-\frac{3}{2} \right)-{2  \over \pi^2}  {\rm Im} \left( {\rm Li}_2(\ri\, \re^\xi)  \right)\lambda + \sum_{k \geq 3}f_{0,k}\lambda^k, \\
	\CF_1(\lambda, \xi)&= -\frac{1}{12}\log \hbar -\frac{1}{12}\log \lambda +\zeta'(-1)+ \sum_{k \geq 1}f_{1,k}\lambda^k, \\
	\CF_g(\lambda, \xi)&=\frac{B_{2g}}{2g(2g-2)}\lambda^{2-2g}+ \sum_{k \geq 1}f_{g,k}\lambda^k, \qquad g \geq 2.
\ea
\ee
In writing the second term in the first line, we used the dilogarithm identity
\be
{\rm Li}_2(z) + {\rm Li}_2\left({1\over z}\right) = -{\pi^2 \over 6}- {1\over 2}  \log^2(-z). 
\ee
In the last line, $B_{2g}$ are Bernoulli numbers. The coefficients $f_{g,k}$ are themselves non-trivial functions of the parameter $\xi$. For $g=0$, one finds, at the very first orders, 
\be
\ba
\label{fgk}
	f_{0,3}&=  \pi^2\frac{1-3 \cosh(2\xi)}{24 \cosh(\xi)},\\
	f_{0,4}&=\pi^4 \frac{-73+68\cosh(2\xi)+45 \cosh(4\xi)}{2304 \cosh^2(\xi)}, \\
	f_{0,5}&=\pi^6\frac{534-203 \cosh(2\xi)-390 \cosh(4\xi)-165 \cosh(6\xi)}{30720 \cosh^3(\xi)},
	\ea
	\ee
	while for $g=1,2$, one finds, 
	\be
	\label{fgkbis}
	\ba
	f_{1,1}&=\pi^2 \frac{-1+3\cosh(2\xi)}{48 \cosh(\xi)}, \\
	f_{1,2}&=\pi^4 \frac{127+4\cosh(2\xi)-27\cosh(4\xi)}{2304 \cosh^2(\xi)}, \\
	f_{1,3}&=\pi^6 \frac{-750-265\cosh(2\xi)+30\cosh(4\xi)+57\cosh(6\xi)}{18432 \cosh^3(\xi)}, \\
	f_{2,1}&=\pi^6 \frac{894+577\cosh(2\xi)+210\cosh(4\xi)+15\cosh(6\xi)}{61440 \cosh^3(\xi)}.
\ea
\ee
These results will be crucial in order to compare the asymptotic evaluation of the fermionic spectral traces, to the predictions of \cite{ghm}. 

\subsection{The exact planar solution}

The $O(2)$ matrix model can be solved exactly in the planar limit \cite{ks,ek}. However, it was noted in \cite{gm} that instead of using the 
specific results for the $O(2)$ case, it is more convenient to first consider the $O(n)$ model for arbitrary $n$, solve it with the powerful techniques of \cite{ek2}, and then 
take the limit $n \rightarrow 2$. 

In order to proceed, we change variables $z=\re^u$ 
in the matrix integral (\ref{zf0-bis}), and we obtain
\begin{align} \label{Zpartitioninzvar} 
Z(N, \hbar)=\frac{\re^{-\frac{\xi}{2 \pi \mg^2}\lambda}}{N!} \int \frac{\rd^N z}{(2\pi)^N} \re^{-\frac{1}{\mg}\sum_{i=1}^N (V^{(0)}(z_i)+\mathcal O(\mg^2))} \frac{\prod_{i < j} (z_i-z_j)^2}{\prod_{i , j} (z_i+z_j)},
\end{align}
where the classical potential (\ref{vo}), when written in terms of $z$, reads
\begin{align}
\label{pp-z}
	V^{(0)}(z)=-\frac{\log(z)}{2\pi}+\frac{{\rm Im \, Li}_2(\ri \, z \re^{\xi})+{\rm Im \, Li}_2(\ri \, z \re^{-\xi})}{\pi^2}.
\end{align} 
To obtain the planar limit it is enough to consider the classical potential in (\ref{Zpartitioninzvar}). 
We assume that we can model the distribution of eigenvalues by a continuous function on a single connected compact support, i.e. 
we assume that we have a one-cut solution. This is a natural assumption, since the potential has a unique minimum at $u=0$ and it has a linearly 
confining behavior (\ref{confinement}). We will 
take the cut along the segment $[a,b] \in \mathbb R_+$. Following the techniques of \cite{ek2} (in the conventions of \cite{gm}) we introduce the auxiliary $G$-functions, 
\begin{align}
	G^{(\nu)}(z)&=-\ri \Big (\re^{\frac{\ri \pi \nu}{2}}G^{(\nu)}_+(z)-\re^{-\frac{\ri \pi \nu}{2}}G^{(\nu)}_+(-z) \Big ), \\
	G^{(1-\nu)}(z)&=-\Big (\re^{\frac{\ri \pi \nu}{2}} g_+(z) G^{(\nu)}_+(z)+\re^{-\frac{\ri \pi \nu}{2}}  g_+(-z)  G^{(\nu)}_+(-z) \Big ),
\end{align}
where
\begin{align}
	G^{(\nu)}_+(z)&=
	\frac{-\ri z}{\sqrt{z^2-a^2}\sqrt{z^2-b^2}} \frac{\vartheta_4(0) \vartheta_1\left(\pi\frac{v-\ri(1-\nu){\rm K'}}{2 {\rm K}}\right)}
	{\vartheta_4 \left(\pi \frac{\ri(1-\nu){\rm K'}}{2\rm K}\right) \vartheta_1\left(\pi \frac{v}{2 \rm K} \right)} \re^{-\frac{\ri \pi (1-\nu) v}{2 {\rm K}}} \qquad  {\rm with}
	\qquad z=a \, {\rm sn}(v), \\ \nonumber
	\\
	g_+(z)&=\frac{\sqrt{z^2-a^2}\sqrt{z^2-b^2}+\frac{z}{e}\sqrt{e^2-a^2}\sqrt{e^2-b^2}}{z^2-e^2}  \qquad  {\rm with}\qquad e=a \, {\rm sn}(\ri(1-\nu){\rm K'}). \\ \nonumber
\end{align}
Here, ${\rm K}$ and ${\rm K'}$ are elliptic integrals of the first kind, and $\vartheta_i(u)$ are Jacobi theta functions. We follow the conventions of \cite{AS} for all the 
elliptic functions and integrals appearing in these formulae. The elliptic modulus $k$ and the nome $q \,\, (=\re^{\ri \pi \tau})$ are given by:
\begin{align}
	k=\frac{a}{b}, \qquad \qquad
	q=\re^{-\pi \frac{\rm K'}{\rm K}}.
\end{align}
Also, the $\nu$ parameter is related to the $n$ of the $O(n)$ model by  
\be
n=2\cos(\pi \nu),
\ee
so that $n \rightarrow 2$ corresponds to $\nu \rightarrow 0$.

Let us denote by $\mathcal C$ the closed contour encircling the cut at $[a,b]$ clockwise. 
The following equations, due to \cite{ek2}, allow to find the 't Hooft parameter $\lambda$ as a function of the end-points of the cut $a,b$:
\begin{align}
	0&=\frac{1}{2 \cos\left(\frac{\pi(1-\nu)}{2}\right)}\oint_{\mathcal C} \frac{\rd z}{2 \pi \ri } {\rd V^{(0)}\over \rd z} G^{(1-\nu)}(z), \\
	\lambda&=\frac{1}{2(1-\cos(\nu \pi)) \cos(\frac{\pi \nu}{2})}\oint_{\mathcal C} \frac{\rd z}{2 \pi \ri } z {\rd V^{(0)} \over \rd z} G^{(\nu)}(z).
\end{align}
The first equation is satisfied if we set $b=1/a$, as expected from the symmetry $u_i \leftrightarrow -u_i$ 
of the matrix integral. So our elliptic modulus is given by $k=a^2$.
The second equation leads, in the limit $\nu \rightarrow 0$, to the equation
\begin{align}
	\lambda={f(\xi,a) \over  \pi^2}.
\end{align}
To determine the function $f(\xi,a)$, we note that the derivative $\partial f(\xi,a)/\partial \xi$ can be computed by deforming the contour and using the residue theorem. Indeed, we have
\begin{align}
	\frac{\partial}{\partial \xi} \left(z {\rd V^{(0)} \over \rd z} \right)=\frac{z}{2\pi^2 \ri}\Big (  \frac{1}{z+\ri \re^\xi} -\frac{1}{z-\ri \re^\xi} - \frac{1}{z+\ri \re^{-\xi}}+\frac{1}{z-\ri \re^{-\xi}}\Big ).
\end{align}
After some calculations, one obtains
\be
\ba
\frac{\partial}{\partial \xi} f(\xi,a)&=\frac{\rm K}{2 \pi \sqrt{(a^2+1)^2+4a^2 \sinh^2(\xi)}}\Biggl\{ -\left( {\rm Z}\left(\arcsin \frac{\re^{\xi}}{ \sqrt{a^2+\re^{2\xi}} }\right) -\frac{\re^{\xi} \sqrt{1+a^2 \re^{2\xi}} }{ \sqrt{a^2+\re^{2\xi}} } \right)^2  \\ & \qquad \qquad 
	+\left( {\rm Z}\left(\arcsin \frac{\re^{-\xi}}{ \sqrt{a^2+\re^{-2\xi}} }\right) -\frac{\re^{-\xi} \sqrt{1+a^2 \re^{-2\xi}} }{ \sqrt{a^2+\re^{-2\xi}} }\right)^2 +2a^2 \sinh(2\xi) \Biggr\},
\ea
\ee
where ${\rm Z}$ is the Jacobi Zeta function. 
The argument of the elliptic functions appearing in this and subsequent expressions is now given by the complementary modulus 
\be
\label{mone}
k_1=\sqrt{1-a^4}.
\ee
Since $f(\xi,a) \rightarrow 0$ when $\xi \rightarrow -\infty$, we can write
\begin{align} 
	f(\xi,a)=\int_{-\infty}^\xi \rd \xi' \frac{\partial}{\partial \xi'} f(\xi',a).
\end{align}
A convenient expression for this integral is in terms of Jacobi theta functions with nome 
\be
\label{q1t1}
q_1=\re^{-\pi \frac{ {\rm K'}}{{\rm K}} }=\re^{\ri \pi \tau_1}. 
\ee
One finds, 
\be
\label{reduced}
\ba
	f(\xi,a)&=\lim_{\Lambda \rightarrow \infty}  \biggl[ \frac{1}{4} \int_{\frac{\pi}{2 \rm K}w(\xi)}^{\frac{\pi}{2\rm K}w(\Lambda)}   \left( \frac{\vartheta_2'}{\vartheta_2} (y )^2 - \frac{\vartheta_1'}{\vartheta_1}(y)^2  \right) 
	\rd y\\ & \qquad \qquad +\frac{\rm K}{2 \pi} \left( \sqrt{(a^2+1)^2+4a^2 \sinh^2(\xi)} 
		-\sqrt{(a^2+1)^2+4a^2 \sinh^2(\Lambda)} \, \right)  \biggr], 
\ea
\ee
where 
\be
w(\xi)={\rm F} \left( \arcsin \frac{\re^\xi}{\sqrt{a^2+\re^{2\xi}}}\right)
\ee
and ${\rm F}$ is the incomplete elliptic integral of the first kind with modulus $k_1$. 
This equation determines the endpoints of the cut as functions of the 't Hooft parameter $\lambda$. 
The planar free energy is then determined by the equation \cite{ek2,gm}, 
\begin{align} \label{F0ofq}
	\frac{\rd^2 \mathcal F_0}{\rd \lambda^2}=-2 \pi \frac{ {\rm K'}}{{\rm K}}=2 \log q_1, 
\end{align}
up to two integration constants, which can be easily fixed by the weak coupling analysis of the previous section.

The exact planar solution makes it possible to explore the dependence of $\CF_0$ on the full moduli space of $\lambda, \xi$. First of all, 
we can reproduce the perturbative results in (\ref{fg-stru}), (\ref{fgk}) by doing a small 't Hooft coupling expansion. When the 't Hooft parameter 
goes to zero, the cut collapses to the minimum of the potential. In the $z$-plane, the endpoint of the cut $a$ goes towards $1$, 
so we can expand in small $k_1=\sqrt{1-a^4}$. In this case, we can use 
the $q_1$-expansions of the theta functions in (\ref{reduced}), and after some calculations we find, 
\be
\label{lambdaofm1} 
\ba
	\lambda&=\frac{1}{64 \pi^2 \cosh(\xi)} k_1^4+\frac{1}{64 \pi^2 \cosh( \xi)} k_1^6+\frac{115+119 \cosh (2\xi)}{16384 \pi^2 \cosh^3 (\xi)} k_1^8
	\\
	&+\frac{51+55 \cosh (2\xi)}{8192 \pi^2 \cosh^3 (\xi)} k_1^{10} +\mathcal O(k_1^{12}).
	\ea
	\ee
Inverting this series and plugging it in (\ref{F0ofq}), we obtain, after integrating twice, 
\be
\label{wF0}
\ba
	\mathcal F_0( \lambda, \xi)&=c_0(\xi)+\left(c_1(\xi)- \frac{\xi}{2\pi} \right) \lambda
	+\frac{\lambda^2}{2} \left( \log \frac{\pi^2 \lambda  \cosh \xi}{4}- \frac{3}{2} \right)+\frac{\pi^2(1-3\cosh(2\xi))}{24 \cosh(\xi)}\lambda^3
	\\ 
	&  +\frac{\pi^4(-73+68 \cosh(2\xi)+45 \cosh(4\xi))}{2304 \cosh^2(\xi)}\lambda^4+\mathcal O(\lambda^5), 
\ea
\ee
where $c_{0,1}(\xi)$ are integration constants, and we added after integration the missing $-\xi \lambda/(2\pi)$ from the prefactor of (\ref{Zpartitioninzvar}). 
This agrees with the perturbative expansion at genus zero from (\ref{fg-stru}), (\ref{fgk}). 

One advantage of the exact solution is that we can also analyze the regime of strong 't Hooft coupling. For this, we do an $S$-transformation in (\ref{reduced}) 
and express our formulae in terms of 
\be
\label{qtt1}
q=\re^{\ri \pi \tau}=\re^{-\ri \pi /\tau_1}, 
\ee
 which is the relevant variable for the large $\lambda$ expansion. We also do a shift in the integration variable to obtain:
\begin{align} \nonumber 
	f(\xi,a)=\lim_{\Lambda \rightarrow \infty}  \Biggl\{ -\frac{\tau}{4} \int_{\frac{\pi \tau}{2}(1-w(\xi)/{\rm K})}^{\frac{\pi \tau}{2}(1-w(\Lambda)/{\rm K})}  \rd y \left( \frac{\vartheta_1'}{\vartheta_1} (y ) - \frac{\vartheta_4'}{\vartheta_4}(y)  \right) \left( \frac{\vartheta_1'}{\vartheta_1} (y ) +\frac{\vartheta_4'}{\vartheta_4}(y) +\frac{4 \ri y}{\pi \tau} \right)   \\
	+\frac{\rm K}{2 \pi} \left( \sqrt{(a^2+1)^2+4a^2 \sinh^2(\xi)}
	-\sqrt{(a^2+1)^2+4a^2 \sinh^2(\Lambda)} \, \right )  \Biggr\},
\end{align}
where the elliptic integrals are still evaluated at $k_1$. As we did for the weak coupling expansion, we expand the integrand in small $q$ and integrate. After some calculations, we obtain
\begin{align} \nonumber
	\lambda&=\frac{1}{8 \pi^3} \log^2\frac{k}{4}-\frac{1}{12\pi}-\frac{\xi^2}{2\pi^3}+\frac{1}{\pi^3}\cosh(2\xi)\left(1-\log\frac{k}{4}\right) \left( \frac{k}{4}\right) \\
	&\qquad \qquad \qquad  +\frac{1}{4\pi^3} \left\{ 4\left(1-\log \frac{k}{4}\right)+3\cosh(4\xi)\left(-1+2\log\frac{k}{4}\right)\right\} \left( \frac{k}{4}\right)^2+\mathcal O(k^{3}).
\end{align}
where we remind that $k=a^2$. This can be inverted to yield the series, 
\begin{align} 
	\frac{k}{4}&=\re^{-2\pi \sqrt{2 \hat \lambda}}+\left(4+\frac{\sqrt{2}}{\pi \sqrt{\hat \lambda}} \right) \cosh(2\xi)\re^{-4\pi \sqrt{2 \hat \lambda}}+\mathcal O\left(\re^{-6\pi \sqrt{2 \hat \lambda}}\right),
\end{align}
where we use the shorthand notation
\begin{align} \label{lambdashift}
	\hat \lambda=\pi \lambda +\frac{1}{12}+\frac{\xi^2}{2\pi^2}.
\end{align}
By using again (\ref{F0ofq}), we finally obtain, 
\be
\label{sce}
\ba
	\mathcal F_0(\lambda, \xi)&=-\frac{\sqrt{2} \, \hat \lambda^{3/2}}{3\pi}+\tilde c_0(\xi)+\left( \tilde c_1(\xi)- \frac{\xi}{2\pi} \right) \lambda-\frac{\cosh(2\xi)}{4 \pi^4} \re^{-2 \pi \sqrt{2 \hat \lambda}}  \\
	&\, \, -\frac{1}{32 \pi^5} \left\{ \left(8 \pi+\frac{4}{\sqrt{2 \hat \lambda}}\right)+\left(\pi+\frac{4}{\sqrt{2 \hat \lambda}}\right) \cosh(4\xi) \right\}\re^{-4 \pi \sqrt{2 \hat \lambda}} +\mathcal O\left(\re^{-6 \pi \sqrt{2 \hat \lambda}}\right), 
\ea
\ee
where $\tilde c_{0,1}(\xi)$ are integration constants that can be fixed in principle from the weak coupling behavior. 
As anticipated in \cite{ghm}, the strong coupling expansion of the free energy displays the $3/2$ scaling typical of theories of M2 branes \cite{kt}, and the coefficient of the leading term agrees with the 
general formula for local del Pezzo Calabi--Yau's found in \cite{ghm}. The expansion (\ref{sce}) is very similar to the expansion of the planar free energy of ABJ(M) theory presented in \cite{dmp}. As 
we will see in the next section, one can in fact recover the result for ABJ(M) theory from (\ref{sce}). Let us also note that, in the case $\xi=0$, the formulae above simplify considerably, and one can write the periods in terms of indefinite integrals of theta functions, 
\be
\label{xi0}
\ba
	\lambda&=-\frac{1}{2\pi \ri} \int \rd \tau_1 \,\vartheta_2(2 \tau_1)^4 \vartheta_3(2 \tau_1)^2, \\
	\frac{\rd \CF_0}{\rd \lambda}&= - \int \rd \tau_1 \, \tau_1 \, \vartheta_2(2 \tau_1)^4 \vartheta_3(2 \tau_1)^2 +c, 
\ea
\ee
where $c$ is again an integration constant. These integrals can be performed in order to obtain an expression which is useful for strong coupling expansions, namely, 
\be
\label{lf-ex-meijer}
\ba
	\lambda&=\frac{1}{4 \pi^4} G_{3,3}^{3,2}\left(\frac{4 k}{(k+1)^2} \left |
\begin{array}{c}
 \frac{1}{2},\frac{1}{2},1 \\
 0,0,0 \\
\end{array}
\right. \right) +C_1 , \\
	\frac{\d \mathcal F_0}{\d \lambda}&= \frac{4k}{16 \pi(k+1)^2} \, _4F_3 \left ( 1,1,\frac{3}{2},\frac{3}{2} ; 2,2,2 ; \frac{4 k}{(k+1)^2}  \right ) +\frac{1}{4\pi } \log \left (-\frac{4 k}{(k+1)^2} \right) +C_2.
\ea
\ee

Another ingredient of the planar solution which can be computed exactly is the density of eigenvalues. Let us first consider the resolvent of the $O(2)$ matrix model (\ref{Zpartitioninzvar}), defined as
\be
\omega(p)= \frac{1}{N} \left\langle \tr {1\over p-M}\right \rangle, 
\ee
where $M$ is the matrix with eigenvalues $z_i$, $i=1, \cdots, N$, and the bracket denotes the normalized vev. We can split this function into its even and odd parts with respect to $p$, 
\be
\omega(p)=\omega_+(p)+\omega_-(p). 
\ee
The planar limit of the even part can be computed by using the formula \cite{ek}
\begin{align} 
	\omega_{+}^0(p)=-\frac{1}{2\lambda} \oint_{\mathcal C} \frac{\rd w}{2\pi \ri} \frac{V'(w) w}{p^2-w^2} \frac{\sqrt{(p^2-a^2)(p^2-\frac{1}{a^2})}}{ \sqrt{(w^2-a^2)(w^2-\frac{1}{a^2})} },
\end{align}
where $V(z)$ is the planar potential (\ref{pp-z}), and $\mathcal C$ is a contour around the cut $[a,1/a]$ anti-clockwise. We find, 
\begin{align}  \nonumber
	\omega_{+}^0(p)&=\frac{a \sqrt{(p^2-a^2)(p^2-\frac{1}{a^2})}}{4 \pi^2 \ri p^2 \lambda}
	 \left\{ \Pi \left(\frac{a^2}{p^2},\arcsin\frac{\ri \re^{\xi}}{a} \Big| a^4 \right) +\Pi \left(\frac{a^2}{p^2},\arcsin\frac{\ri \re^{-\xi}}{a} \Big | a^4 \right)  - L(p,a) \right\} \\
	&\qquad \qquad \qquad \qquad \qquad \qquad \qquad \qquad +\frac{1}{4 \pi^2  p \lambda} \left( \arctan\frac{\re^\xi}{p} - \arctan \re^\xi p \right) ,
\end{align}
where $\Pi$ is the elliptic integral of the third kind, and 
\begin{align}
 	L(p,a)= \lim_{\Lambda \rightarrow \infty}  \Pi \left(\frac{a^2}{p^2},\ri \Lambda \Big | a^4 \right) = \frac{-\ri p^2}{a^2(a^2-p^2)} \left [ \Pi \left (1-\frac{p^2}{a^2} \right| 1-\frac{1}{a^4} \right)- {\rm K'} \left(\frac{1}{a^4} \Big ) \right].
 \end{align}
When $z \in [a,1/a] $, the eigenvalue density is given by the discontinuity equation 
 \begin{align}
 	\rho(z)&=\frac{1}{\ri \pi} \Big ( \omega_{+}^0(z+\ri 0) - \omega_{+}^0(z-\ri 0) \Big  ) \\
		&=\frac{a \sqrt{(z^2-a^2)(\frac{1}{a^2}-z^2)} }{2 \pi^3 \ri z^2 \lambda} \left\{  \Pi \left(\frac{a^2}{z^2},\arcsin\frac{\ri \re^{\xi}}{a} \Big | a^4 \right) +\Pi \left(\frac{a^2}{z^2},\arcsin\frac{\ri \re^{-\xi}}{a} \Big |  a^4 \right)  - L(z,a) \right\}.
 \end{align}
 %
\begin{figure}[h!]
\center
	\includegraphics[scale=0.55]{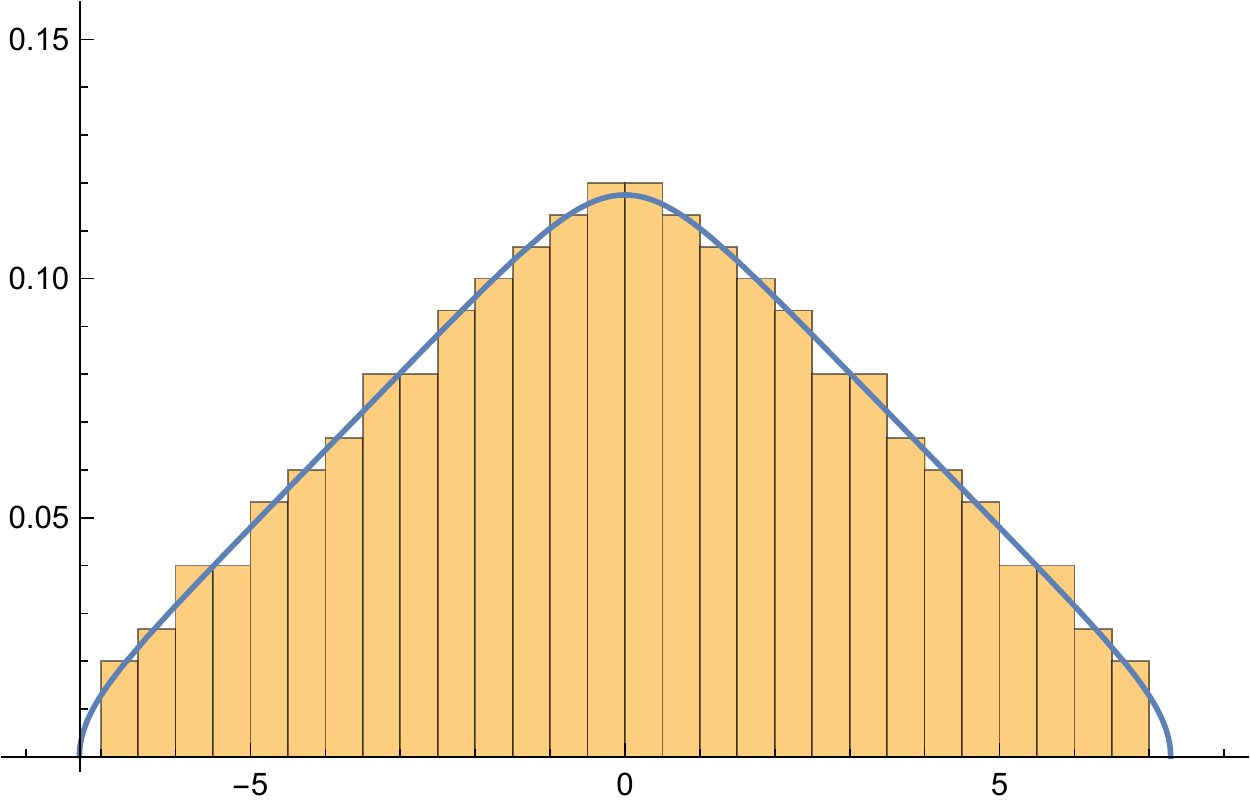} \quad \includegraphics[scale=0.55]{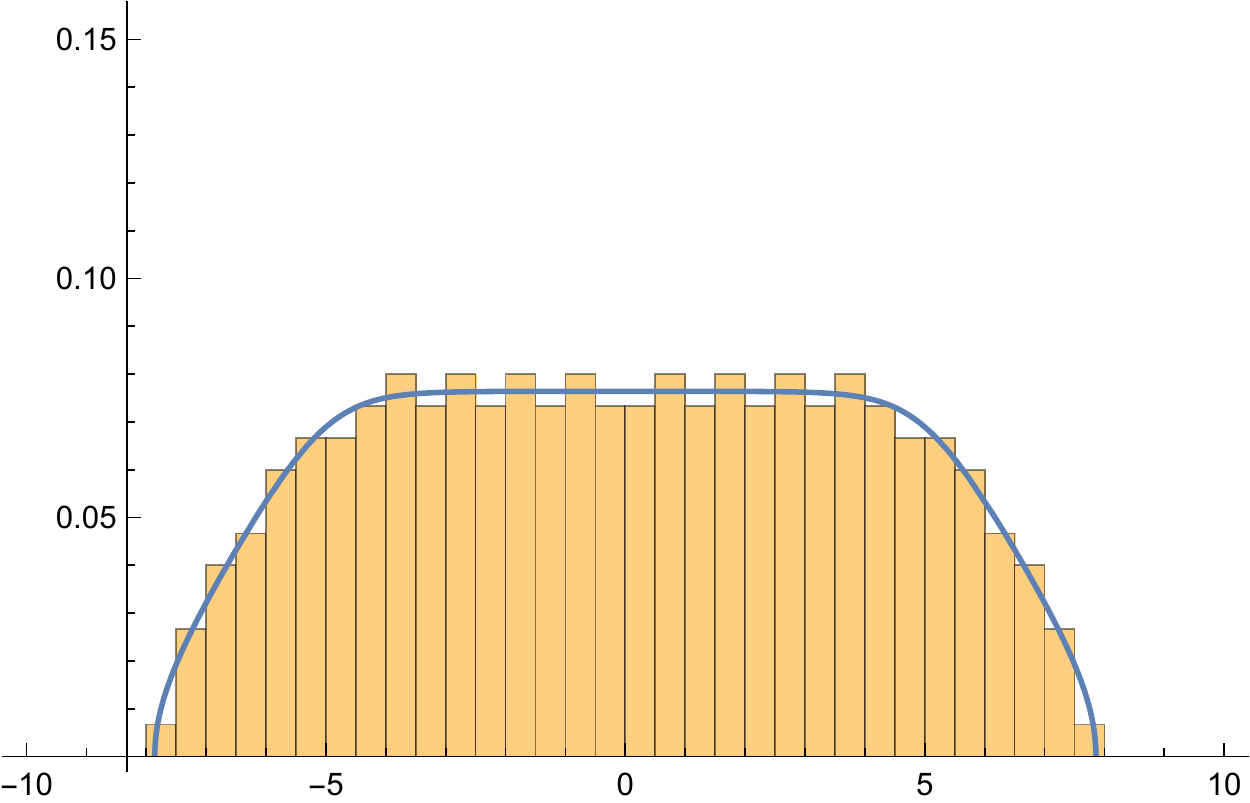} 
	\caption{Left: eigenvalue density $\rho(u)$ for $\lambda=1$, $\xi=0$ against a histogram showing the numerical density of $N=300$ relaxed eigenvalues. Right: same plot with $\lambda=3/4$ and $\xi=5$.}
\label{density-f}
\end{figure}
This expression can be checked by doing a numerical simulation of 300 eigenvalues relaxed into a configuration which approximately minimizes 
the effective action of the matrix model, as in \cite{hkpt}. The 
results are shown in \figref{density-f}, after going back to the initial  $u$ variable, so that $\rho(u)=\rho(z(u))\rd z/\rd u$, with $z(u)=\re^u$.

\subsection{Relation to the ABJ(M) matrix model}
\label{abj-sec}

In \cite{kwy}, a matrix model computing the partition function of ABJ(M) theory \cite{abjm,abj} on $\IS^3$ was derived, by using localization. This model turns out to 
be closely related to the topological string on local $\IP^1 \times \IP^1$. In the case of ABJM theory, this was first noted and exploited in \cite{mp, dmp} in the context of the 't Hooft expansion, and then 
in \cite{hmmo,kallen-m} for the M-theory expansion (which involves as well the refined topological string in the NS limit). The generalization to ABJ theory was done in 
\cite{abj-moriyama,honda-o}. Since the matrix model (\ref{zf0}) gives a non-perturbative completion of the partition function on this geometry, it is natural to 
wonder whether it is related to the ABJ(M) matrix model. In fact, one can recover the ABJ(M) matrix model from (\ref{zf0}) provided one considers 
{\it complex} values of the parameter $m_{\IF_0}$. Note that the operator (\ref{f0}) is no longer self-adjoint in this case, and in addition one has to be careful 
with the resulting multi-valued structure, since 
the integral kernel depends on the logarithm of $m_{\IF_0}$.

Let us then set 
\be
\label{m-M}
\log\, m_{\IF_0}=\ri \hbar -2\pi \ri M,  \qquad M \in \IZ_{\ge 0}. 
\ee
Here, the integer $M$ will be identified with the difference between the ranks of the two Chern--Simons theories in ABJ theory \cite{abj}. 
This relationship is the one suggested by the explicit results of \cite{abj-moriyama, honda-o}. In these papers, the grand potential of ABJ(M) theory is written in terms of the topological 
string on local $\IF_0$. 
If we now use the explicit expression (\ref{rho-ex}) for the integral kernel, we find
\begin{align} \nonumber
	\rho_{\rm \mathbb F_0}=\re^{-\ri \hbar/4+\ri \pi M/2} \re^{\sqrt{\pi \hbar} \,\mq  /2}
	 \frac{{\rm \Phi}_{ \sqrt {\hbar/\pi}}( \mq+\frac{\ri}{2} M\sqrt{\pi/\hbar})}{ {\rm \Phi}_{\sqrt {\hbar/\pi}}( \mq -\frac{\ri}{2}M \sqrt{\pi/\hbar}) }
	  \frac{1}{2 \cosh(\sqrt{\pi \hbar} \, \map)}  \qquad \qquad
	  \\ 
	 \times  \frac{{\rm \Phi}_{ \sqrt {\hbar/\pi}}( \mq-\frac{\ri}{2} M\sqrt{\pi/\hbar}+\frac{\ri}{2} \sqrt{\hbar/\pi})}{ {\rm \Phi}_{\sqrt {\hbar/\pi}}( \mq +\frac{\ri}{2} M\sqrt{\pi/\hbar}-\frac{\ri}{2} \sqrt{\hbar/\pi}) } \re^{\sqrt{\pi \hbar} \, \mq   /2}.
\end{align}
Due to the form of the arguments, we can use the functional equations for the quantum dilogarithm, (\ref{eq.bshift}) and (\ref{eq.tbshift}), and we obtain
\be
\ba
	\rho_{\rm \mathbb F_0}=\re^{-\ri \hbar/4+\ri \pi M/2} \re^{\sqrt{\pi \hbar} \, \mq  /2}
	\left( \prod_{s=\frac{-M+1}{2}}^\frac{M-1}{2} \frac{1}{1+\re^{2\pi (\mq \sqrt{\pi/\hbar}+\ri \pi s / \hbar )}} \right)
	\frac{1}{2 \cosh(\sqrt{\pi \hbar} \, \map )}  \qquad \qquad
	  \\ 
	\times \left( \prod_{s=\frac{-M+1}{2}}^\frac{M-1}{2} (1+\re^{2\pi (\mq \sqrt{\pi/\hbar}+\ri \pi s / \hbar +\ri/2)}) \right) \frac{1}{1+(-1)^M\re^{2 \sqrt{\pi \hbar} \, \mq }}
	   \re^{\sqrt{\pi \hbar} \, \mq  /2}.
\ea
\ee
To make contact with ABJ(M) theory, let us define
\be
\label{khbar}
k={\hbar \over \pi}
\ee
 and let us introduce the variables
 \be
 \mathsf{u}=2\pi \sqrt{k} \, \mq, \qquad {\mathsf{v}}=2\pi \sqrt{k} \, \map, 
 \ee
so that $[\mathsf{u}, \mathsf{v}]=2 \pi \ri k$. In these new variables, and after a similarity transformation, we find, 
\be
\label{f0-abj}
	A \,  \rho_{\rm \mathbb F_0} \,  A^{-1}=\re^{-\ri \pi k/4+\ri \pi M/2} \rho_{\rm ABJ(M)}, 
	\ee
	where
	\be
	\rho_{\rm ABJ(M)}=\frac{1}{2 \cosh(\mathsf{v}/ 2)} \frac{1}{\re^{ \frac{\mathsf{u}}{2} }+(-1)^M \re^{-\frac{\mathsf{u}}{2} }}  \prod_{s=\frac{-M+1}{2}}^\frac{M-1}{2} \tanh \Big ( \frac{\mathsf{u} +2\pi \ri s}{2 k} \Big ) 
\ee
is, up to a similarity transformation, the operator appearing in the Fermi gas formulation of ABJM theory \cite{mp} and of ABJ theory \cite{ahs,honda,honda-o}. Since the phase appearing in (\ref{f0-abj}) 
is the same one appearing in the relation between $\IF_0$ and $\IF_2$, we also conclude that, 
\be
\rho_{\IF_2}= \rho_{\rm ABJ(M)}, 
\ee
up to a combination of unitary and similarity transformations. The dictionary between the parameters is (\ref{khbar}) and 
\be
m_{\IF_2}= 2 \cos\left({\pi k \over 2}-\pi M \right). 
\ee
In particular, the spectral traces of the kernel of the ABJ(M) matrix model can be obtained from the traces of the $\IF_2$ operator. 
This can be easily tested for ABJM theory, in the case $k=2$, $M=0$, by using the expressions (\ref{trf2}). 
The relevant value of the mass parameter is $m_{\IF_2}=-2$, which is a branch point for the functions in (\ref{trf2}). 
However, the traces at this point are well-defined and one finds the correct values \cite{hmo}
\be
\tr \, \rho_{\rm ABJM}\big|_{k=2}={1\over 8}, \qquad \tr \,  \rho^2_{\rm ABJM}\big|_{k=2}={1\over 64}-{1\over 16 \pi^2}. 
\ee
We also find that, when $k=2$ and $M=1$, which is the maximally supersymmetric ABJ theory, the theory is equivalent (at the level of spectral traces) to the maximally supersymmetric case of local $\IF_2$ with 
$m_{\IF_2}=2$, or equivalently of local $\IF_0$ with $m_{\IF_0}=1$. 

The exact results for the planar solution found in the previous section can be used to re-derive the exact planar solution 
of the ABJ(M) matrix model, first obtained in \cite{mp-abjm,dmp}. 
Indeed, due to (\ref{m-M}), the exact planar free energy of ABJ(M) theory can be obtained from the general formulae obtained above by setting
\be
\xi=\frac{\ri \pi}{2}-\frac{\ri \pi^2 M}{\hbar}. 
\ee
As a check, note that the shifted variable (\ref{lambdashift}) becomes 
\be
\label{abj-shift} 
\hat \lambda=\pi \lambda -\frac{1}{2}\Big ( B^2 -\frac{1}{4}\Big )-\frac{1}{24},
	\ee
where
\be
B=\frac{1}{2}-{M\over k}
\ee
has to be identified as the $B$ field of ABJ theory. The shift (\ref{abj-shift}) is precisely the one found in \cite{dmp}. In addition, 
the strong coupling expansion (\ref{sce}) becomes, 
\be 
\label{abj-f}
\ba 
	{\mathcal F_0}(\lambda, \beta)&=-\frac{\sqrt{2} \, \hat \lambda^{3/2}}{3\pi}+\tilde c_0+\Big ( \tilde c_1- \frac{\ri}{4} \Big )\lambda+
	\frac{\beta+\beta^{-1}}{8 \pi^4} \re^{-2 \pi \sqrt{2 \hat \lambda}} \\
	\nonumber
	& \qquad \qquad
		-\frac{1}{4 \pi^4} \left\{ \frac{1}{16}(\beta^2+16+\beta^{-2})+\frac{1}{4\pi \sqrt{2 \hat \lambda}}(\beta+\beta^{-1})^2 \right\} 
		\re^{-4 \pi \sqrt{2 \hat \lambda}} +\mathcal O\left(\re^{-6 \pi \sqrt{2 \hat \lambda}}\right), 
		\ea
		\ee
where
\be
\beta=\re^{-2 \pi \ri M/k}. 
\ee
The function in (\ref{abj-f}) is precisely $ -1/(4 \pi^4)$ times the planar free energy $F_0^{\rm ABJ}$ obtained in \cite{dmp}. This overall factor is due to our different conventions 
for the string coupling constant. 

It should be noted however that the perturbative and weak coupling expansion worked out for the matrix model (\ref{zf0-bis}) can not be used for ABJM theory in the form presented above. 
For ABJM theory, $M=0$, so that $\xi = \ri \pi/2$, and the expansions (\ref{fgk}), (\ref{fgkbis}) and (\ref{wF0}) diverge. This is not a problem of our exact solution, 
but rather a breakdown of the Gaussian approximation. 
The reason is that, when considering the particular limit of ABJM theory, the ``classical" potential (\ref{vo}) is no longer a perturbed Gaussian, since it is exactly given 
by the r.h.s. of (\ref{confinement}), and in particular it is not smooth at $u=0$. One can however still obtain the correct weak coupling 
expansion from the exact planar solution, and one obtains, 
\begin{align} 
	 \lambda=\frac{1}{8\pi^2}m_1+\frac{1}{16\pi^2}m_1^2+\frac{65}{1536 \pi^2}m_1^3+ \mathcal O(m_1^4),
\end{align}
where $m_1$ is given by (\ref{mone}), as well as 
\begin{align} 
	\mathcal F_0(\lambda)&=c_0+c_1 \lambda + \left( \log \left(\frac{\pi^2 \lambda}{2}\right) -\frac{3}{2} \right)\lambda^2-\frac{\pi^4 }{9}\lambda^4+\frac{283 \pi^8}{5400}\lambda^6-\frac{961 \pi^{12}}{19845}\lambda^{8} +\mathcal O(\lambda^{10}), 
\end{align}
which is precisely (up to an overall factor $ -1/(4 \pi^4)$) the expression found in \cite{dmp}. In addition, one finds the relations
\be
\label{xipi}
\ba
	\lambda&=-\frac{1}{2\pi \ri} \int \rd \tau_1 \,\vartheta_3(2 \tau_1)^4 \vartheta_2(2 \tau_1)^2, \\
	\frac{\rd \CF_0}{\rd \lambda}&= - \int \rd \tau_1 \, \tau_1 \, \vartheta_3(2 \tau_1)^4 \vartheta_2(2 \tau_1)^2 +\tilde c, 
\ea
\ee
which are obtained from (\ref{xi0}) by exchanging $\vartheta_2$ with $\vartheta_3$. This can be also integrated explicitly, as in (\ref{lf-ex-meijer}), and the result is in precise agreement with the result 
of \cite{dmp}.

\sectiono{Comparing the matrix model to the topological string}

\subsection{Predictions from the spectral theory/mirror symmetry correspondence}

The conjecture of \cite{ghm} gives a very precise prediction for the 't Hooft expansion (\ref{free-expansion}) of the fermionic traces of the 
trace class operators obtained by quantizing mirror curves. We will now summarize some of the results of \cite{ghm}, specialized to the case of interest, namely local $\IP^1\times \IP^1$ 
(see also \cite{km,mz} for other summaries of the main results of \cite{ghm}). According to the conjecture of \cite{ghm}, the basic 
quantity determining the spectral properties of the operator $\rho_{\IF_0}$ is the modified grand potential $J(\mu, m_{\IF_0}, \hbar)$. This function depends 
on the ``chemical potential" $\mu$, which is related to the ``fugacity" $\kappa$ entering in (\ref{f-det}) as 
\be
\kappa=\re^\mu, 
\ee
as well as on the mass parameter $m_{\IF_0}$ appearing in (\ref{f0}). The modified grand potential is determined by the enumerative geometry of the CY. 
We first need a dictionary between the parameters $\mu$, $m_{\IF_0}$, and the parameters appearing in the enumerative geometry of local $\IP^1 \times \IP^1$. This CY has a 
``diagonal" K\"ahler parameter $T$, which is related to $\mu$ by 
\be
\label{genkal}
T=2 \mu_{\rm eff}.  
\ee
Here, the ``effective" $\mu$ parameter is determined by the so-called quantum mirror map of \cite{acdkv} (see \cite{ghm} for the notation),
\be
\label{mueff}
\mu_{\rm eff}= \mu- {1\over 2} \sum_{\ell \ge 1}  \widehat a_\ell (\hbar) \re^{-2 \ell \mu}. 
\ee
In this paper we will not need the explicit expression of this quantum mirror map, since it is not relevant in the 't Hooft limit we will focus on. 
In addition, there is a K\"ahler parameter $T_m$ associated to the mass parameter $m_{\IF_0}$. Geometrically it 
measures, roughly speaking, the difference in sizes between the two spheres in local $\IP^1 \times \IP^1$, and one 
has the relation
\be
T_m= -\log  m_{\IF_0}. 
\ee
The modified grand potential has the form, 
 \be
 \label{jx-masses}
 J(\mu, m_{\IF_0}, \hbar) = J^{({\rm p})} (\mu_{\rm eff}, m_{\IF_0}, \hbar) + J_{\rm M2} (\mu_{\rm eff}, m_{\IF_0}, \hbar) + J_{\rm WS}(\mu_{\rm eff}, m_{\IF_0}, \hbar). 
 \ee
Here, $J^{({\rm p})} (\mu, m_{\IF_0}, \hbar) $ is the perturbative part, which is a cubic polynomial in $\mu$:
 \be
 \label{jp}
 J^{({\rm p})} (\mu, m_{\IF_0}, \hbar)={2 \over 3 \pi \hbar } \mu^3 -{\log\, m_{\IF_0} \over 2 \pi \hbar} \mu^2 +\left( {\pi \over 3\hbar}-{\hbar  \over 12 \pi} \right) \mu + A(m_{\IF_0}, \hbar). 
 \ee
 When $m_{\IF_0}=1$ one recovers the expression presented in \cite{ghm}. When $m_{\IF_0}\not=1$, the part which depends on $\mu$ can be obtained in a relatively 
 straightforward way by working out the classical grand potential \cite{yhu,gkmr}, or by using the result for local $\IF_2$ and the dictionary between this model and 
 local $\IF_0$. A precise expression for the function $A(m_{\IF_0}, \hbar)$ has been obtained by Y. Hatsuda \cite{yhu}. His expression exploits the relationship between 
 ABJ theory and topological string theory on local $\IP^1\times \IP^1$ discussed in section \ref{abj-sec}. It is given by, 
 \be
 \label{amh}
	A(m_{\IF_0},\hbar)=\frac{\log^3 m_{\IF_0}}{48 \pi \hbar}-\frac{\log m_{\IF_0}}{4}\Big ( \frac{\pi}{3 \hbar}-\frac{\hbar}{12 \pi}\Big )+A_{\rm c} \left( {\hbar \over \pi} \right)
	-F_{\rm CS}\left( {\hbar \over \pi},M\right). 
\ee
Let us spell out the details of this formula. The function $A_{\rm c}(k)$ was introduced in \cite{mp} in the Fermi gas approach to ABJM theory. 
It was determined explicitly in \cite{hanada} and further simplified in \cite{ho}. It reads, 
\be
\label{ak}
A_{\rm c}(k)= \frac{2\zeta(3)}{\pi^2 k}\left(1-\frac{k^3}{16}\right)
+\frac{k^2}{\pi^2} \int_0^\infty \frac{x}{\re^{k x}-1}\log(1-\re^{-2x})\rd x.
\ee
In (\ref{amh}), $F_{\rm CS}(k,M)$ is an analytic continuation of the Chern--Simons free energy on the three-sphere for gauge group $U(M)$ and level $k$, 
\be
F_{\rm CS} (k,M)= \log \, Z_{\rm CS}(k,M), 
\ee
where $M$ is related to the parameters of our problem as
\be
\label{Mk}
M= {\hbar + \ri \log m_{\IF_0}\over 2 \pi}. 
\ee
Note that this is precisely the relation (\ref{m-M}) used in section \ref{abj-sec}. As it is well-known, the Chern--Simons 
partition function for integer $M$ is given by \cite{witten}
\be
Z_{\rm CS}(k,M)= k^{-M/2} \prod_{j=1}^M \left( 2 \sin {\pi j \over k} \right)^{M-j}, 
\ee
but in view of (\ref{Mk}) we have to extend it to arbitrary complex $M$. This can be done in various equivalent 
ways \cite{yhu,kota,ruben}, but in this paper we will not need the 
precise form of this extension.

The ``membrane" part of the potential $J_{\rm M2} (\mu_{\rm eff}, m_{\IF_0}, \hbar)$ appearing in (\ref{jx-masses}) will not be relevant for our purposes. It is fully determined by the 
 refined BPS invariants of the topological string in this CY background, see \cite{ghm,mz} for details. Finally, the worldsheet part of the 
 modified grand potential is 
\be
\label{wsj}
J_{\rm WS}(\mu_{\rm eff}, m_{\IF_0}, \hbar)= \sum_{g\ge 0} \sum_{v=1}^{\infty} \sum_{ {\bf d}} n^{\bf d}_g {1\over v} \left( 2 \sin {2 \pi^2 v\over \hbar} \right)^{2g-2} 
\re^{- {2 \pi \over \hbar } v {\bf d} \cdot {\bf T}}, 
\ee
where 
\be
 {\bf T}=(T, T_m),
 \ee
$n^{\bf d}_g$ are the Gopakumar--Vafa invariants \cite{gv2} of local $\IP^1\times \IP^1$, and ${\bf d}= (d_1, d_2)$ are two non-negative integers. 

One of the consequences of the conjecture of \cite{ghm} is that the fermionic spectral traces $Z_{\IF_0}(N, \hbar)$ can be obtained as 
integral transforms of the modified grand potential,
\be
\label{zn-airy}
Z_{\IF_0}(N, \hbar)={1\over 2 \pi \ri} \int_\CC \re^{J(\mu,m_{\IF_0}, \hbar) - N \mu} \rd \mu, 
\ee
where the contour $\CC$ goes from $\re^{-\ri \pi/3} \infty$ to $\re^{\ri \pi/3} \infty$ (so that the integral is absolutely convergent). 
The formula (\ref{zn-airy}) leads to a precise prediction for the 't Hooft limit of the fermionic spectral traces. Note that, in order to keep 
the dependence on both K\"ahler parameters, we have to take 
a limit in which $\log m_{\IF_0}$ scales with $\hbar$, as we required in (\ref{mscaling}). We then consider the 't Hooft limit of $J(\mu,m_{\IF_0}, \hbar)$, in which  
\be
\label{j-limit}
\mu \rightarrow \infty, \qquad m_{\IF_0}\rightarrow \infty, \qquad \hbar \rightarrow \infty, 
\ee
and
\be
{\mu \over \hbar}= \zeta \, \, \, {\rm fixed}, \qquad {\pi \over 2 \hbar} \log m_{\IF_0}= \xi \, \, \, {\rm fixed}. 
\ee
The parameter $\xi$ was introduced in (\ref{xidef}). We will express often the results in terms of
\be
\label{mxi}
m=\re^{4 \xi}, 
\ee
which corresponds to the mass parameter appearing in the standard topological string free energies. Indeed, with this definition, one has that
\be
-\log(m)= {2 \pi \over \hbar} T_m. 
\ee

In the 't Hooft limit, the membrane part of the grand potential in (\ref{jx-masses}) goes to zero, and $\mu_{\rm eff} \rightarrow \mu$. The remaining ingredients have non-trivial 
't Hooft-like expansions. The expansion of $A(m_{\IF_0}, \hbar)$ can be easily worked out. The function $A_{\rm c}(k)$ has the large $k$ expansion \cite{hanada}:
\be
A_{\rm c}(k)=-{k^2\over 8 \pi^2} \zeta(3)+ {1\over 2} \log(2) + 2 \zeta'(-1) +{1\over 6} \log \left({\pi \over 2k}\right) + \sum_{g\ge 2} \left({2 \pi \over k}\right)^{2g-2}  4^g (-1)^{g-1} c_g, 
\ee
where
\be
c_g={B_{2g} B_{2g-2} \over (4g) (2g-2)(2g-2)!}.
\ee
On the other hand, in the limit we are considering, $M \rightarrow \infty$ but
\be
\label{cs-thooft}
{2 \pi^2 \ri \over \hbar} M = \pi \ri - 2\xi 
\ee
is fixed. This is the standard 't Hooft expansion of $F_{\rm CS}(\hbar/\pi,M)$, worked out at all genus in \cite{gv}, and with 't Hooft parameter (\ref{cs-thooft}). 
One then finds an expansion of the form, 
\be
A(m_{\IF_0}, \hbar)= \sum_{g\ge 0} A_g(\xi)\hbar^{2-2g}, 
\ee
where $A_1(\xi)$ includes as well a logarithmic dependence on $\hbar$, and 
\be
\label{tHooftA}
\ba
A_0(\xi)&=\frac{\zeta(3)-2 {\rm Li}_3(-\re^{2\xi})}{8 \pi^4}, \\
A_1(\xi)&=-\frac{\xi}{6}+\frac{1}{12}\log \left(16  \pi^2 \cosh \xi  \right)-\frac{1}{12}\log \hbar+\zeta'(-1), \\
A_g(\xi)&=(2 \pi^2)^{2g-2}  (-1)^{g-1}  \left\{ (4^g-2)  c_g-{B_{2g} \over 2g(2g-2)!} {\rm Li}_{3-2g}\left(-\re^{2\xi} \right) \right\}.
\ea
\ee
It follows from this expression that
\be
\label{ag0}
A_g(0)=2 (-1)^{g-1} \left( 4 \pi^2 \right)^{2g-2} c_g  \left(3-2^{3-2g}\right), \qquad g\ge 2, 
\ee
in agreement with the result presented in \cite{ghm} for $m_{\IF_0}=1$, 
\be
\label{am0}
A\left(m_{\IF_0}=1, \hbar \right)= {3\over 2} A_{\rm c}\left( {\hbar \over \pi}\right)-A_{\rm c}\left( {2 \hbar \over \pi} \right). 
\ee

One concludes that, in the 't Hooft limit (\ref{j-limit}), the modified grand potential has the asymptotic expansion, 
\be
J^{\text{'t Hooft}}(\zeta, \xi, \hbar) = \sum_{g=0}^\infty J_g(\zeta, \xi) \hbar^{2-2g}, 
\ee
where
\be
\label{jg-par}
\ba
J_0(\zeta, \xi)&=\frac{2}{3\pi}\zeta^3-\frac{\log m}{4\pi^2}\zeta^2-\frac{1}{12 \pi}\zeta+A_0(\xi)+\frac{1}{16\pi^4}F_0^{\rm inst}(t,m), \\
J_1(\zeta, \xi)&=\frac{\pi}{3}\zeta+A_1(\xi)+F^{\rm inst}_{1} (t,m), \\
J_g(\zeta, \xi)&= A_g(\xi) + (4 \pi^2)^{2g-2} F_g^{\rm inst}\left( t, m \right), \qquad g\ge 2. 
\ea
\ee
Here, we have introduced the variable
\be
\label{tr}
t= 4 \pi \zeta, 
\ee
and $F^{\rm inst}_g(t, m)$ is the worldsheet instanton part of the standard genus $g$ topological string free energy. In order to obtain the 't Hooft expansion of the fermionic trace and 
make contact with (\ref{thooft-f0}), we have to calculate the integral in (\ref{zn-airy}), by doing a saddle-point 
expansion for $\hbar$ large. Let us denote by 
\be
\label{conj-exp}
\sum_{g\ge 0} \CF_g^{\IF_0} (\lambda, m) \hbar^{2-2g}
\ee
the asymptotic expansion of the logarithm of the integral in (\ref{zn-airy}). At leading order, one finds 
\be
\label{l-jo}
\lambda= {\partial J_0( \zeta, m) \over \partial \zeta}, 
\ee
which determines the 't Hooft parameter $\lambda$ as a function of $\zeta$, and conversely, $\zeta$ as a function of $\lambda$. The genus zero 
free energy $\CF_0^{\IF_0} ( \lambda, m)$ is then given by a Legendre transform, 
\be
\label{genus-zero}
\CF_0^{\IF_0} (\lambda, m)= J_0(\zeta(\lambda), m)- \lambda \zeta(\lambda). 
\ee
In particular 
\be
\label{dual-per}
{\partial \CF_0^{\IF_0} \over \partial \lambda}= - \zeta. 
\ee
The next-to-leading order correction to the saddle-point, $\CF^{\IF_0}_1(\lambda, m)$, is given by, 
\be
\label{gones}
\CF_1^{\IF_0} (\lambda, m)= J_1(\zeta(\lambda), m)- {1\over 2} \log \left( 2 \pi {\partial^2 J_0 \over \partial \zeta^2} \right).
\ee
The higher order corrections can be computed systematically by using the results of \cite{abk} (already exploited in this context in 
\cite{mz}): in the saddle point approximation, the integral in (\ref{zn-airy}) 
implements a transformation from the large radius frame, appropriate to $J_g(\zeta, m)$, to the conifold frame. Therefore, the functions $ \CF_g^{\IF_0} (\lambda, m)$ 
appearing in (\ref{conj-exp}) are the topological 
string free energies of local $\IP^1 \times \IP^1$ in the conifold frame. 
According to the conjecture of \cite{ghm}, they should be equal to the matrix model free energies appearing in (\ref{thooft-f0})
\be
\label{conj-eq}
\CF_g(\lambda, m)=\CF_g^{\IF_0} (\lambda, m), \qquad g\ge 0. 
\ee
This was tested in detail for local $\IP^2$ and for local $\IF_2$ (for a fixed valued of $m_{\IF_2}$) in \cite{mz}. We will devote the rest of this paper to an explicit 
verification of (\ref{conj-eq}), and in the next section we will compute the r.h.s. of (\ref{conj-eq}) by using standard techniques in topological string theory.

\subsection{Topological strings on local $\IP^1\times \IP^1$}

Let us review some basic facts about the special geometry of local $\IP^1 \times \IP^1$. Since this has two two-cycles, we can regard it as a two parameter model, and its mirror has then 
two complex moduli $z_1$, $z_2$. However, it has been known for some time that $m=z_1/z_2$ does not receive quantum corrections, therefore it should be rather regarded 
as a parameter (see for example \cite{hkp}). We will then have a complex modulus, $z=z_2$, and a ``mass" parameter $m$. 
The periods will be obtained as solutions to a single Picard--Fuchs (PF) equation corresponding to the operator \cite{hkp}: 
\begin{align} \nonumber
	\mathcal L =\Big ( 8(1-m)^2 z^2-4(1+m)z+\frac{1}{2} \Big ) \theta^3 + \Big ( 16(1-m)^2 z^2-4(1+m)z \Big ) \theta^2 \\ \label{singlePF}
	\qquad +\Big ( 6(1-m)^2 z^2-(1+m)z \Big ) \theta, 
\end{align}
where
\be
\theta=z{\rd \over \rd z}. 
\ee
This is the form of the operator which is appropriate for the large radius point $z=0$. 
As usual in local mirror symmetry, there will be a constant solution $1$, a logarithmic solution 
\be
g_1(z)= \log(z) + \sigma_1(z), 
\ee
and a double logarithmic solution, 
\be
 g_2(z)=\log^2(z)+2 \log(z) \sigma_1(z)+\sigma_2(z). 
\ee
In these equations, $\sigma_{1,2}(z)$ are power series around $z=0$, whose coefficients depend on $m$. The very first orders read, 
\be
\ba
\label{sigmas}
\sigma_1(z)&=2(m^{\frac{1}{2}}+m^{-\frac{1}{2}})m^{\frac{1}{2}}z+3 \Big( (m+m^{-1})+4 \Big)m z^2\\
&+\frac{20}{3}\Big ((m^{\frac{3}{2}}+m^{-\frac{3}{2}})+9(m^{\frac{1}{2}}+m^{-\frac{1}{2}}) \Big)m^{\frac{3}{2}}z^3+\mathcal O(z^4),\\
\sigma_2(z)&=4(m^{\frac{1}{2}}+m^{-\frac{1}{2}})m^{\frac{1}{2}}z+\Big ( 13(m+m^{-1})+40 \Big)m z^2\\
&+\frac{8}{9}\Big (41(m^{\frac{3}{2}}+m^{-\frac{3}{2}})+279(m^{\frac{1}{2}}+m^{-\frac{1}{2}}) \Big)m^{\frac{3}{2}}z^3+\mathcal O(z^4)
\ea
\ee
Let us now consider the following linear combinations of the basic periods, 
\begin{align} \label{periodALR}
	\Pi^{\rm (lr)}_A(z)&=\Big ( 0 \qquad 1 \qquad 0 \Big ) 
		 \left (
	  \begin{array}{ l}
  	1 \\
	g_1(z)\\
	g_2(z)
  \end{array} \right ),
  \\ 
  \label{periodBLR}
  \Pi^{\rm (lr)}_B(z)&=\Big ( 0  \qquad \frac{\log m}{2} \qquad \frac{1}{2} \Big ) 
		 \left (
	  \begin{array}{ l}
  	1 \\
	g_1(z)\\
	g_2(z)
  \end{array} \right ).
\end{align}
The first A-period determines the flat coordinate $t$ through the mirror map, while the second, B-period determines the genus zero free energy $F_0(t,m)$ at large radius, 
\begin{align}
	-t=\Pi^{\rm (lr)}_A, \qquad \qquad \frac{\d F_0}{\d t}=\Pi^{\rm (lr)}_B.
\end{align}
After integration, we get, 
\begin{align}
\label{F0inLR}
	F_0(t,m)&= \frac{t^3}{6}-\frac{\log m}{4}t^2-2(m^{\frac{1}{2}}+m^{-\frac{1}{2}}) m^{\frac{1}{2}}\re^{-t}-\frac{1}{4} \Big ( (m+m^{-1})+16\Big )m \re^{-2t} 
	\\ 
	& \qquad \qquad
	-\frac{2}{27}\Big ( (m^{\frac{3}{2}}+m^{-\frac{3}{2}}) +81(m^{\frac{1}{2}}+m^{-\frac{1}{2}})\Big )m^{\frac{3}{2}} \re^{-3t}+\mathcal O(\re^{-4t}). 
\end{align}
Equivalently, one can obtain the same information from the equations
\begin{align} \label{exactLR1}
	\frac{\d t}{\d z}=-\frac{2}{\pi z \sqrt{1-4(\sqrt{m}+1)^2z}} {\rm K} \left( \frac{16 \sqrt{m}z}{4(\sqrt{m}+1)^2z-1} \right), \\ \label{exactLR2}
	\frac{\d^2 F_0}{\d t \d z}=-\frac{2}{ z \sqrt{1-4(\sqrt{m}-1)^2z}} {\rm K} \left( \frac{4(\sqrt{m}+1)^2z-1}{4(\sqrt{m}-1)^2z-1} \right), 
\end{align}
which can be obtained from the results of \cite{bt} for local $\IF_2$, together with the dictionary relating the moduli of local $\IF_2$ to those of local $\IF_0$. 

We now analyze the theory near the conifold locus given by the vanishing of the discriminant:
\be
\Delta=1-8(m+1)z+16(m-1)^2 z^2 =\Big( 4(1+\sqrt{m})^2z-1 \Big ) \Big ( 4(1-\sqrt{m})^2z-1 \Big ).
\ee
Note that there are two different branches of the conifold locus, related to the two square roots of $m$. For each value of 
$m$, we have a different conifold point in each of the branches of 
the conifold locus, and we have to analyze the topological string near an arbitrary point, as a function of $m$. For $m=1$, the topological string 
near the corresponding conifold point at $z=1/16$ has 
been analyzed in \cite{hkr,dmp,mz}. We will pick for convenience the branch of positive roots, and introduce the local 
variable, 
\be 
\label{conifoldvars}
y=1-4(1+{\sqrt{m}})^2 z,  
\ee
which vanishes at the conifold point 
\be
\label{z-coni}
z_c= \frac{1}{4(1+ \sqrt{m})^2}.
\ee
In the variables appropriate to the conifold point, the PF operator becomes
\be \ba
	\mathcal{\tilde L}=& \, \, 4(y-1)^2 \Big ( y(\mu-1)^2+4\mu \Big ) \theta_{y}^3 \\ 
	&+4(y-1) \Big (2y^2(\mu-1)^2+y(1+\mu)^2 +8\mu \Big ) \theta_{y}^2 \\
	&+\Big ( 3y^3(\mu-1)^2+4y^2\mu+y(\mu^2-6\mu+1)+16\mu \Big ) \theta_{y}.
\ea
\ee
There is a basis of solutions given by a constant solution $1$, a vanishing solution 
\be
f_1(y)=y+\mathcal O(y^2)
\ee
 and a logarithmic solution 
 \be
 f_2(y)=\log(y) f_1(y)+s(y), \qquad s(y)=y+\mathcal O(y^2). 
 \ee
It is easy to solve for $f_1(y)$ and $s(y)$ as power series in $y$ with $m$-dependent coefficients:
\be
\ba
f_1(y)&=y-\frac{\cosh(2\xi)-11}{16}y^2+\frac{9\cosh(4 \xi)-124 \cosh(2 \xi)+827}{1536}y^3+\mathcal O(y^4), \\
s(y)&=y-\frac{7\cosh(2\xi) -45}{32}y^2+\frac{27\cosh(4 \xi)- 380 \cosh(2 \xi) +1561}{1152}y^3+\mathcal O(y^4), 
\ea
\ee
where we expressed the results in terms of the variable $\xi$, related to $m$ by (\ref{mxi}). The analytic continuation of the large radius periods to the conifold point 
must be a linear combination of the two solutions $f_1(y)$, $f_2(y)$ found above. By expanding the exact results (\ref{exactLR1})-(\ref{exactLR2}) around the 
conifold locus, one finds
\begin{align} 
   \label{combi2}
  \Pi^{\rm (lr)}_A(z)&=\Big (C_1  \qquad \frac{\cosh \xi}{\pi} \Big (\log \Big ( \frac{\cosh^2\xi}{16} \Big  )-2\Big ) \qquad \frac{ \cosh \xi}{\pi}  \Big ) 
		 \left (
	  \begin{array}{ l}
  	1 \\
	f_1(y)\\
	f_2(y)
  \end{array} \right ),\\
  \label{combi1}
	\Pi^{\rm (lr)}_B(z)&=\Big ( C_2 \qquad  \pi \cosh \xi \qquad 0 \Big ) 
		 \left (
	  \begin{array}{ l}
  	1 \\
	f_1(y)\\
	f_2(y)
  \end{array} \right ),
\end{align}
where $C_1$, $C_2$ are {\it a priori} $\xi$-dependent constants which we do not know how to evaluate analytically. These constants are given by the 
values of the large radius periods at the conifold point, i.e. 
\be
C_1=  \Pi^{\rm (lr)}_A(z_c), \qquad C_2=\Pi^{\rm (lr)}_B(z_c). 
\ee
The constant $C_1$ can be computed analytically, and we present this computation in Appendix \ref{con-A}. The constant $C_2$ 
can be calculated numerically, by evaluating the series (\ref{sigmas}) at the conifold point 
(where the series still converges). However, as we will see in the next section, the value of these constants is {\it predicted} by the conjecture of \cite{ghm}, and we will find a precise agreement with the 
analytical and numerical evaluations of $C_{1,2}$, respectively. 

The above results determine the genus zero free energy at large radius $F_0(t, m)$. The genus one free energy can be obtained, for example, 
from the result for local $\IF_2$ in \cite{bt}, by using the map of moduli. One finds, 
\be
\label{F1LR}
F_{1}(t,m)=-\frac{1}{12}\log \Big (m^{\frac{1}{2}}z^7(16(m-1)^2z^2-8(m+1)z+1) \Big )-\frac{1}{2}\log\left( -\frac{\d t}{\d z} \right), 
\ee
with the large radius expansion 
\be
\label{F1ex}
F_1(t,m)= -{1\over 24} \log (m) +{t\over 12} -{1\over 6}(1+m) \re^{-t}+ \cdots
\ee
The higher genus free energies near the conifold point 
can be obtained by integrating the holomorphic anomaly equation. A systematic computation has only been performed at $m=1$ \cite{hkr,dmp}, 
but it will provide us with useful tests, as we will see in the next section. 

\subsection{Comparison}
Let us now compare the results from the matrix model/spectral theory side, with the predictions from the conjecture of \cite{ghm}. We start with the genus zero free energy. 
By using the expansion (\ref{F0inLR}), we can write 
\be
\label{J0ofT}
J_0 \left(\zeta=\frac{t}{4\pi},\xi \right)=\frac{1}{16 \pi^4}\left( F_0(t, m)-\frac{\pi^2}{3}t+16 \pi^4 A_0(\xi) \right), 
\ee
where $t$ is related to $\zeta$ by (\ref{tr}) and it is the standard K\"ahler parameter of the geometry. We also recall that $m$ is related to $\xi$ by (\ref{mxi}). 
The 't Hooft parameter is given by (\ref{l-jo}), which reads in this case, 
\be
4 \pi^3 \lambda=\Pi_B^{\rm (lr)}-\frac{\pi^2}{3}. 
\ee
The form of the matrix model expansion (\ref{fg-stru}) indicates that $\lambda$ must be a vanishing period at the conifold point, i.e. 
\be
\Pi_B^{(\rm lr)}(z=z_c)=C_2=\frac{\pi^2}{3}.
\ee
This gives a prediction for the value of the constant $C_2$, which we have verified by evaluating this constant numerically. This test involves doing a 
high precision numerical sum of the large radius expansion of $\Pi_B^{(\rm lr)}(z=z_c)$, for different values of $m$.  

We should note that, as far as we know, the constants that have to be added to the $B$-periods in order to obtain a vanishing period at the 
conifold point are not known {\it a priori}, and they 
have to be determined on a case-by-case basis, often numerically (see \cite{gl,dlo} for some examples.) 
According to the conjecture of \cite{ghm}, this constant is obtained from the terms in the modified grand potential which are 
linear in the chemical potential and next-to-leading in $\hbar$. On the other hand, it follows 
from the general construction in \cite{ghm} that these terms are in turn 
determined by the linear terms in the moduli appearing in the large radius free energy $F_1^{\rm NS}$. 
Therefore, we have the following consequence of the conjecture of \cite{ghm}: in a toric CY threefold, 
{\it the constant terms which appear in the vanishing periods at the conifold are determined by 
the coefficients of the linear terms in $F_1^{\rm NS}$}. This is of course 
also the case for the example of local $\IP^2$ studied in \cite{mz}, and provides an intriguing link between refined 
genus one free energies and the special geometry of the conifold point. 

Let us now proceed to compute $\CF^{\IF_0}_0$ from (\ref{dual-per}), which reads 
\be
\frac{\d \mathcal F^{\IF_0}_0}{\d \lambda}=\frac{1}{4\pi}\Pi_A^{\rm (lr)}.
\ee
This can be integrated to find, up to an integration constant, 
\be
\ba
\label{f0-conifold}
\mathcal F^{\IF_0}_0(\lambda, \xi)&= \frac{\lambda^2}{2} \Big (\log \Big (\frac{\pi^2 \lambda \cosh \xi}{4} \Big )-\frac{3}{2} \Big )+\frac{C_1(\xi)}{4\pi}\lambda  +  \pi^2\frac{1-3 \cosh(2\xi)}{24 \cosh(\xi)} \lambda^3\\
	&  + \pi^4 \frac{-73+68\cosh(2\xi)+45 \cosh(4\xi)}{2304 \cosh^2(\xi)} \lambda^4 \\
	&+
	\pi^6\frac{534-203 \cosh(2\xi)-390 \cosh(4\xi)-165 \cosh(6\xi)}{30720 \cosh^3(\xi)}\lambda^5 +\mathcal O(\lambda^6). 
\ea
\ee
The $\lambda$-independent function $C_1(\xi)$ appearing here is the one appearing in (\ref{combi1}). This result agrees with the 
results in (\ref{fg-stru}), (\ref{fgk}), and (\ref{wF0}) obtained in the matrix model, provided that
\be
\label{c1-answer}
C_1(\xi)=-{8 \over \pi} {\rm Im} \left( {\rm Li}_2(\ri \re^\xi) \right).
\ee
Note that, as in the examples of \cite{mz}, the r.h.s. of (\ref{c1-answer}) is a {\it prediction} of the conjecture of \cite{ghm} 
on the value of the $A$-period at the conifold point. This prediction comes from the explicit form of the potential (\ref{vo}), which is in turn 
determined by the explicit form of the integral kernel (\ref{f0ker}). As shown in Appendix \ref{con-A}, the value (\ref{c1-answer}) agrees precisely with the analytic evaluation of the 
A-period at the conifold. 

Finally, we note that, according to the result (\ref{fg-stru}) in the matrix model, the integration constant in $\mathcal F^{\IF_0}_0(\lambda,m)$ should vanish. This implies that
\begin{align}
0=\mathcal F^{\IF_0}_0(\lambda=0, m)=\Big ( J_0(\zeta(\lambda), m)-\lambda \, \zeta(\lambda,m) \Big ) \Big |_{\lambda=0}.
\end{align}
Using (\ref{J0ofT}), this yields the following relation between the function $A_0(\xi)$ in (\ref{tHooftA}) and the value of the genus zero free energy at the conifold point:
\be
\label{A0numerical}
A_0(\xi)=\frac{1}{16\pi^4} \Big ( F_0(-C_1(\xi), \xi)+\frac{\pi^2}{3} C_1(\xi) \Big ),
\ee
where we used that $t(z_c)=-C_1(\xi)$. This is yet another remarkable consequence of the conjecture of \cite{ghm} for the special geometry of the conifold point, and we have 
verified it numerically with high precision. To give a flavor of the validity of (\ref{A0numerical}), in \figref{A0-fig} we show the value of $A_0(\xi)$, as given in (\ref{tHooftA}), against a numerical evaluation of the r.h.s. of (\ref{A0numerical}) for 
some values of $\xi$. 
\begin{figure}[h]
\center
\includegraphics[scale=0.7]{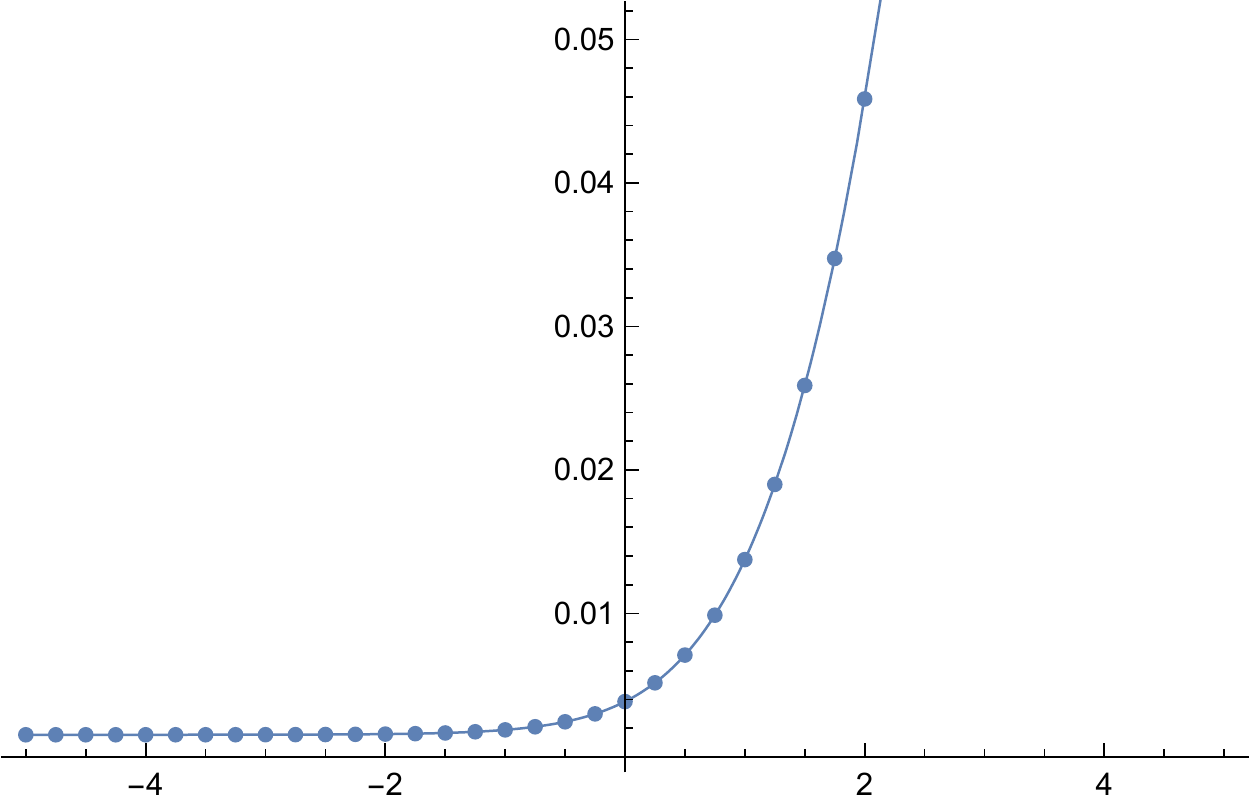}
\caption{The continuous line shows the exact function $A_0(\xi)$, as given in (\ref{tHooftA}), 
while the dots are numerical evaluations of the r.h.s. of (\ref{A0numerical}) for some values of $\xi$.
}
\label{A0-fig}
\end{figure}

We conclude that, at the level of the genus zero free energy, there is a remarkable agreement between the result obtained from the matrix model (i.e. from the spectral theory side) 
and the predictions of \cite{ghm} based on topological string theory. In particular, one can use the results of the matrix model/spectral theory side to obtain non-trivial information about the conifold 
theory which is not available otherwise (as far as we know): the analytic values of the periods and the genus zero free energy at the conifold point.

Let us now consider the genus one free energy. Note that
\be
J_1(\zeta, \xi)=\frac{\pi}{3}\zeta+A_1(\xi)+F^{\rm inst}_{1}(t, \xi)=A_1(\xi)+{\xi \over 6}+ F_{1}(t, \xi), 
\ee
where we used (\ref{F1LR}) and the expansion (\ref{F1ex}). If we now take into account (\ref{gones}) and the equation 
\be
\frac{\d^2 J_0 (\zeta)}{\d \zeta^2}=4\pi\frac{\d  \lambda}{\d t},
\ee
we obtain 
\be
\ba
\mathcal F_1^{\IF_0}(\lambda) &= A_1(\xi)+{\xi \over 6}-\frac{1}{24}\log m-\frac{1}{12}\log \Big (m^{\frac{1}{2}}z^7(\lambda)(16(m-1)^2z^2(\lambda)-8(m+1)z(\lambda)+1) \Big )\\ 
& \qquad +\frac{1}{2}\log \Big ( -\frac{1}{8\pi^2} \frac{\d z(\lambda)}{\d \lambda} \Big ).
\ea
\ee
By using the explicit expression for $A_1(\xi)$ in (\ref{tHooftA}), we find that the small $\lambda$ expansion of this function is, 
\be
\ba
	\mathcal F_1^{\IF_0}(\lambda) &= -\frac{1}{12}\log \lambda -\frac{1}{12}\log \hbar+\zeta'(-1) \\
	& \qquad +\pi^2\frac{-1+3 \cosh(2\xi)}{48 \cosh^2(\xi)}\lambda +\pi^4\frac{127+4\cosh(2\xi)-27\cosh(4\xi)}{2304 \cosh^2(\xi)}\lambda^2+\CO(\lambda^3),
\ea
\ee
  which is in precise agreement with what was found in (\ref{fg-stru}), (\ref{fgkbis}). 
  
There are some non-trivial tests that can be done at higher genus, in the case $\xi=0$, by using the results in \cite{dmp,hkr}. 
Let us first recall that the topological string free energies in the conifold frame, when expanded around the conifold point in terms of a vanishing period, 
have a universal critical behavior characterized by a pole of order $2g-2$, for $g\ge 2$ \cite{gvcone}. It was 
then pointed out in \cite{hk} that the full expansion satisfies a ``gap condition," i.e. after this pole, 
the rest of the expansion is regular and it starts at zeroth order in the vanishing period. This has been exploited to constrain 
solutions to the holomorphic anomaly equations. However, in the matrix model free energies, as one can see in (\ref{fg-stru}), 
the expansion around the conifold fulfills a  ``strong" gap condition, in the sense that the expansion in $\lambda$ after the pole starts at {\it first} order in $\lambda$ (and 
not at zeroth order). In contrast, the conventional topological string free energies satisfy only a ``weak" gap condition. In practice, this has the following consequence. 
Let us consider the instanton part of the large radius, genus $g$ free energies $F_g^{\rm inst}(t,m)$, and let us perform a symplectic transformation 
to the conifold frame. The ``weak" gap condition of \cite{hk} implies that the expansion of the resulting quantities around the conifold point is of the form, 
\be
\label{bg-constant}
{B_{2g} \over 2g(2g-2)} t_c^{2-2g}+ b_g(\xi)+ \CO(t_c), 
\ee
where $t_c= 4 \pi^2 \lambda$ is a vanishing period at the conifold\footnote{We are considering just the instanton part of the large radius free energies, so we are not including the 
constant map contribution to these amplitudes to zero. Note however that adding this contribution does not lead in general to a strong gap condition at the conifold. In other words, $-b_g$ is not 
the constant map contribution at large radius.}. Then, it follows from the last line in (\ref{jg-par}) that
\be
\CF_g (\lambda, \xi)= {B_{2g} \over 2g(2g-2)} \lambda^{2-2g}+  (4 \pi^2)^{2g-2} b_g(\xi)+ A_g(\xi)+ \CO(\lambda). 
\ee
Therefore, consistency with the expansion (\ref{fg-stru}), which satisfies a strong gap condition, requires that
\be
\label{bg-ag}
b_g(\xi)=-{A_g(\xi) \over (4 \pi^2)^{2g-2}}, \qquad g\ge 2. 
\ee
This can be regarded as yet  another prediction of spectral theory for the topological string, since the coefficients $A_g(\xi)$ 
have been fixed by consistency with studies of the spectrum. For $\xi=0$, the constants $b_g(0)$ can be computed systematically from the 
holomorphic anomaly equations \cite{hkr, dmp}. One finds, for the very first genera, 
\be
b_2(0)=-{1\over 1152}, \qquad  b_3(0)={23\over 5806080}, \qquad b_4(0)=-{19\over 278691840}, 
\ee
and by using (\ref{ag0}), one verifies that (\ref{bg-ag}) is indeed 
satisfied. 

Finally, we note that the genus two free energy in the 
conifold frame is given by \cite{hkr,dmp}, 
\be
F_2^{\rm inst}\left(t_c, \xi=0\right)= -\frac{1}{240 t_c^2}-\frac{1}{1152}+\frac{53 t_c}{122880}-\frac{2221 t_c^2}{14745600}+\cdots 
\ee
The third term in this expansion agrees with the coefficient $f_{2,1}$ in (\ref{fgkbis}), for $\xi=0$ after taking into account the overall 
factor $(4 \pi^2)^2$ in (\ref{jg-par}). We conclude that the 't Hooft expansion of the fermionic traces, as calculated by the matrix model, is in perfect agreement with 
the predictions of \cite{ghm} (and with the result of \cite{yhu} for the function 
$A (m_{\IF_0}, \hbar)$).

\sectiono{Conclusions and open problems}

In this paper we have found an explicit expression for the integral kernel of the trace class operator associated to the mirror curve of local $\IF_0$, for arbitrary values of the mass 
parameter. This makes it possible to obtain a matrix model computing the fermionic spectral traces of this operator. This model turns out to be an $O(2)$ model, which can be 
exactly solved in the planar limit. Using this matrix model, we have verified in detail that the fermionic spectral traces of (\ref{rho-ex}) provide a non-perturbative definition of the topological 
string on this geometry, in the sense that their asymptotic 't Hooft expansion is given by the genus expansion of the topological string. This provides yet another test of the conjecture in \cite{ghm}. 
In particular, our calculation checks the conjectural form of the quantum-mechanical instanton corrections to the spectral problem proposed in \cite{ghm}. 

There are various obvious problems raised by our results. It would be interesting to improve our checks by calculating higher genus amplitudes 
directly in the matrix model, although this type of calculations are not simple for $O(n)$ models. 
Even at genus zero, it would be interesting to have an analytic proof that the function $\CF_0$ obtained in the matrix model agrees exactly with the genus zero 
free energy of the topological string $\CF_0^{\IF_0}$. Another 
obvious open problem is to extend this type of calculations to other geometries, like for example local $\CB_n$, where $\CB_n$ is the blow-up of $\IP^2$ at $n$ points. 
To do this, we would need an explicit form for the integral kernels of the corresponding trace class operators. It would be also interesting to 
find exponentially small corrections to the 't Hooft expansion 
of the matrix model studied here, in order to construct a 
trans-series expansion of the matrix model free energy, in the spirit of \cite{mmnp} (see \cite{rrr} for a recent, detailed 
case study of trans-series in the quartic matrix model). This could then be compared to the predictions of \cite{ghm} and/or to the trans-series construction 
of \cite{cesv1,cesv2}. 

Another research direction concerns the field theory limit of the model analyzed in this paper. It can be explicitly shown 
that the spectral problem for the operator (\ref{f0}) has a double-scaling limit in which one recovers 
the quantum spectral curve of $SU(2)$ Toda given in \cite{gp}. This corresponds to the field theory limit of the topological string, which is pure $N=2$ Yang--Mills theory \cite{kkv}. 
According to \cite{ns}, the NS limit of the instanton partition function of \cite{nekrasov} 
provides a quantization condition for this spectral problem. This can be verified by using the perturbative WKB approach \cite{mirmor}, but there are also non-perturbative corrections 
(see \cite{krefl,dunne,kpt} for different perspectives on this issue). It would be interesting to analyze this field theory limit by using the tools introduced here. 

Finally, as we have explained, the matrix model in this paper generalizes the ABJ(M) matrix model, and in particular extends it to arbitrary values of $M$. 
This is due to the fact that the dependence on $M$ is through the mass parameter $m_{\IF_0}$, as shown in (\ref{m-M}). In contrast, in the existing matrix models for ABJ theory \cite{kwy,ahs}, 
$M$ has to be in principle a positive integer. This might be useful in order to relate ABJ theory to higher spin theories \cite{cmsy,hhos}.

\section*{Acknowledgements}
We would like to thank David Cimasoni, Santiago Codesido, Ricardo Couso-Santamar\'\i a, Alba Grassi, Jie Gu, Yasuyuki Hatsuda, Albrecht Klemm, 
Kazumi Okuyama, Jonas Reuter and Ricardo Schiappa 
for useful discussions and correspondence. We are particularly thankful to Ricardo Schiappa for a detailed reading of the draft. 
This work is supported in part by the Fonds National Suisse, 
subsidies 200020- 141329, 200021-156995 and 200020-141329, and by the NCCR 51NF40-141869 ``The Mathematics of Physics" (SwissMAP).

\appendix
\sectiono{The quantum dilogarithm}
\label{app-qd}
The quantum dilogarithm $\fad(x)$ is defined by \cite{faddeev,fk,fkv}
\begin{equation}\label{fad}
\fad(x)
=\frac{(\re^{2 \pi \mathsf{b} (x+c_{\mathsf{b}})};q)_\infty}{
(\re^{2 \pi \mathsf{b}^{-1} (x-c_{\mathsf{b}})};\tq)_\infty} \,,
\end{equation}
where
\be
\label{qq}
q=\re^{2 \pi \im \mathsf{b}^2}, \qquad 
\tq=\re^{-2 \pi \im \mathsf{b}^{-2}}, \qquad 
\mathrm{Im}(\mathsf{b}^2) >0
\ee
and 
\be
c_\mb= {\im \over 2} \left(\mb+\mb^{-1}\right). 
\ee
An integral representation in the strip $|\mathrm{Im} z| < |\mathrm{Im} \, c_{\mathsf{b}}|$ is given by 
\be
\fad(x)=\exp \left( \int_{\mathbb{R}+\im\epsilon}
\frac{\re^{-2\im xz}}{4\sinh(z\mathsf{b})\sinh(z\mathsf{b}^{-1})}
{\operatorname{d}\!z  \over z} \right).
\ee
Remarkably, this function admits 
an extension to all values of $\mathsf{b}$ with 
$\mathsf{b}^2\not\in\mathbb{R}_{\le 0}$. $\fad(x)$ is 
a meromorphic function of $x$ with
\be
\text{poles:} \,\,\, c_{\mathsf{b}} + \im \IN \mathsf{b} + \im \IN \mathsf{b}^{-1},
\qquad
\text{zeros:} \,\, -c_{\mathsf{b}} - \im \IN \mathsf{b} - \im\IN \mathsf{b}^{-1} \,.
\ee
The functional equation
\be
\fad(x) \fad(-x)=\re^{\pi \im x^2} \fad(0)^2, 
\qquad
\fad(0)=\left(\frac{q}{\tilde q}\right)^{\frac{1}{48}} =\re^{\pi\im\left(\mathsf{b}^2+\mathsf{b}^{-2}\right)/24}, 
\ee
allows us to move $\fad(x)$ from the denominator to the numerator. In addition, when $\mb$ is either real or on the unit circle, we have 
the unitarity relation
\be
\label{unit-fad}
{\overline{\fad(x)}}={1\over \fad\left( \overline x\right)}. 
\ee

The asymptotics of the 
quantum dilogarithm are given by \cite{ak}
\begin{equation}
\label{eq.as}
\fad(x) \sim \begin{cases}
\fad(0)^2\re^{\pi \im x^2} & \text{when} \quad \Re(x) \gg 0, \\
1             & \text{when} \quad \Re(x) \ll 0.
\end{cases}
\end{equation}

The quantum dilogarithm is a quasi-periodic function. Explicitly,
it satisfies the equations
\begin{subequations}
\begin{align}
\label{eq.bshift}
\frac{\fad(x+c_{\mathsf{b}}+\im \mathsf{b})}{
\fad(x+c_{\mathsf{b}})} 
&= \frac{1}{1-q \re^{2 \pi \mathsf{b} x}} 
\\ 
\label{eq.tbshift}
\frac{\fad(x+c_{\mathsf{b}}+\im\mathsf{b}^{-1})}{
\fad(x+c_{\mathsf{b}})} &= 
\frac{1}{1-\tq^{-1} \re^{2 \pi \mathsf{b}^{-1} x}} \,.
\end{align}
\end{subequations}

When $\mb$ is small, we can use the folllowing asymptotic expansion (see for example \cite{ak,mcintosh}), 
\be\label{as-fd}
\log \fad\left( {x \over 2 \pi \mb} \right) \sim  \sum_{k=0}^\infty \left( 2 \pi \im \mb^2 \right)^{2k-1} {B_{2k}(1/2) \over (2k)!} {\rm Li}_{2-2k}(-\re^x), 
\ee
where $B_{2k}(z)$ is the Bernoulli polynomial. 

\sectiono{The A-period at the conifold point} 
\label{con-A}
In this short Appendix we compute the A-period $\Pi_A^{({\rm lr})}(z)$ at the conifold point $z=z_c$, for arbitrary values of $m$ (or, equivalently, of $\xi$). 
The starting point of this calculation is the integral 
\be
\label{i-dimer}
\CI= {1\over (2 \pi \ri)^2} \int_{\IS^1 \times \IS^1} \log P(z,w) {\rd z \over z} {\rd w \over w}, 
\ee
where
\be
P(z,w)= 2 (x^2+ y^2) + x^2 (z+ z^{-1}) +2 y^2 (w+ w^{-1}). 
\ee
Note that $P(z,w)$ is essentially the polynomial defining the mirror curve to local $\IF_0$, and the integral $\CI$ is the logarithmic Mahler measure of this polynomial. Let us define 
\be
z_1= {x^4 \over 4(x^2+ y^2)^2}, \qquad z_2= {y^4 \over 4(x^2+ y^2)^2}. 
\ee
By expanding $\log P(z,w)$ in power series in $z_{1,2}$, we find 
\be
\CI= \log\left( 2 (x^2+ y^2) \right) -\sum_{n=1}^\infty  \sum_{2k+ 2l=n} {\Gamma(2k+2l) \over \Gamma(1+k)^2 \Gamma(1+l)^2}
z_1^k z_2^l.
\ee
If we identify the variables $z_{1,2}$ with the moduli of local $\IF_0$, we have that 
\be
m={z_1 \over z_2}= \left( {x\over y} \right)^4, 
\ee
and we finally obtain
\be
\label{i-first}
\CI= 2 \log y -{1\over 2} \Pi_A ^{({\rm lr})}(z_c), 
\ee
where $z_c$ is given in (\ref{z-coni}). 

On the other hand, the integral (\ref{i-dimer}) was explicitly computed by Kasteleyn in section 3 of 
\cite{kasteleyn}, in the analysis of the dimer model on the bipartite square lattice on the torus (see 
\cite{david} for a nice summary of the subject.) By writing it as 
\be
\CI= {1\over \pi^2} \int_0^\pi \int_0^\pi \log \left( 4 (x^2 \cos^2 \omega + y^2 \cos^2 \omega') \right) \rd \omega \rd \omega', 
\ee
one can first perform the integral w.r.t. $\omega'$, compute the remaining integral as a power series in $x/y=\re^\xi$, and resum the resulting series in terms of the dilogarithm. One finds, 
\be
\CI= 2\log y + {4 \over \pi} {\rm Im} \left( {\rm Li}_2\left(\ri \re^{\xi} \right) \right). 
\ee
By comparing this to (\ref{i-first}), we conclude that 
\be
\Pi_A ^{({\rm lr})}(z_c)=-{8 \over \pi} \left( {\rm Li}_2\left(\ri \re^{\xi} \right) \right), 
\ee
which is precisely what we obtained from (\ref{c1-answer}).


\begin{thebibliography}{99}
\bibliographystyle{plain}

   
\bibitem{ghm}
 A.~Grassi, Y.~Hatsuda and M.~Mari\~no, ``Topological Strings from Quantum Mechanics,''
  arXiv:1410.3382 [hep-th].

  \bibitem{adkmv}
 M.~Aganagic, R.~Dijkgraaf, A.~Klemm, M.~Mari\~no and C.~Vafa, ``Topological strings and integrable hierarchies,''
  Commun.\ Math.\ Phys.\  {\bf 261}, 451 (2006)
  [hep-th/0312085].
    
    \bibitem{ns}
   N.~A.~Nekrasov and S.~L.~Shatashvili, ``Quantization of Integrable Systems and Four Dimensional Gauge Theories,''
  arXiv:0908.4052 [hep-th].
 
 \bibitem{acdkv}
M.~Aganagic, M.~C.~N.~Cheng, R.~Dijkgraaf, D.~Krefl and C.~Vafa, ``Quantum Geometry of Refined Topological Strings,''
  JHEP {\bf 1211}, 019 (2012)
  [arXiv:1105.0630 [hep-th]].

 \bibitem{mirmor}
   A.~Mironov and A.~Morozov, ``Nekrasov Functions and Exact Bohr-Zommerfeld Integrals,''
  JHEP {\bf 1004}, 040 (2010)
  [arXiv:0910.5670 [hep-th]].

  \bibitem{mp}
 M.~Mari\~no and P.~Putrov, ``ABJM theory as a Fermi gas,''
  J.\ Stat.\ Mech.\  {\bf 1203}, P03001 (2012)
  [arXiv:1110.4066 [hep-th]].

     \bibitem{hmo} 
  Y.~Hatsuda, S.~Moriyama and K.~Okuyama, ``Instanton Effects in ABJM Theory from Fermi Gas Approach,''
  JHEP {\bf 1301}, 158 (2013)
  [arXiv:1211.1251 [hep-th]].

\bibitem{hmo2}
  Y.~Hatsuda, S.~Moriyama and K.~Okuyama, ``Instanton Bound States in ABJM Theory,''
  JHEP {\bf 1305}, 054 (2013)
  [arXiv:1301.5184 [hep-th]].
  
\bibitem{hmmo}
Y.~Hatsuda, M.~Mari\~no, S.~Moriyama and K.~Okuyama, ``Non-perturbative effects and the refined topological string,''
  JHEP {\bf 1409}, 168 (2014)
  [arXiv:1306.1734 [hep-th]].

    \bibitem{kallen-m}
 J.~Kallen and M.~Mari\~no, ``Instanton effects and quantum spectral curves,''
  arXiv:1308.6485 [hep-th].
   
    \bibitem{cgm}
 S.~Codesido, A.~Grassi and M.~Mari–o, ``Exact results in $ \mathcal{N}=8 $ Chern-Simons-matter theories and quantum geometry,''
  JHEP {\bf 1507}, 011 (2015)
  [arXiv:1409.1799 [hep-th]].

\bibitem{hw}
 M.~x.~Huang and X.~f.~Wang, ``Topological Strings and Quantum Spectral Problems,''
  JHEP {\bf 1409}, 150 (2014)
  [arXiv:1406.6178 [hep-th]].
  
\bibitem{mz}
 M.~Mari\~no and S.~Zakany, ``Matrix models from operators and topological strings,''
  arXiv:1502.02958 [hep-th].
  
   \bibitem{km}
  R.~Kashaev and M.~Mari\~no, ``Operators from mirror curves and the quantum dilogarithm,''
  arXiv:1501.01014 [hep-th].
  
     \bibitem{kostov}
 I.~K.~Kostov, ``Exact solution of the six vertex model on a random lattice,''
  Nucl.\ Phys.\ B {\bf 575}, 513 (2000)
  [hep-th/9911023].
  
\bibitem{gkmr}
 J.~Gu, A.~Klemm, M.~Mari\~no and J.~Reuter, ``Exact solutions to quantum spectral curves by topological string theory,''
  JHEP {\bf 1510}, 025 (2015) [arXiv:1506.09176 [hep-th]].
  
  \bibitem{ek}
 B.~Eynard and C.~Kristjansen, ``Exact solution of the $O(n)$ model on a random lattice,''
  Nucl.\ Phys.\ B {\bf 455}, 577 (1995)
  [hep-th/9506193].
  
  \bibitem{ek2}
 B.~Eynard and C.~Kristjansen, ``More on the exact solution of the $O(n)$ model on a random lattice and an investigation of the case $|n| > 2$,''
  Nucl.\ Phys.\ B {\bf 466}, 463 (1996)
  [hep-th/9512052].
 
       \bibitem{gm}
 A.~Grassi and M.~Mari\~no, ``M-theoretic matrix models,'' JHEP {\bf 1502}, 115 (2015)
  [arXiv:1403.4276 [hep-th]].
  
   \bibitem{ikp}
   A.~Iqbal and A.~K.~Kashani-Poor, ``Instanton counting and Chern-Simons theory,''
  Adv.\ Theor.\ Math.\ Phys.\  {\bf 7}, no. 3, 457 (2003) [hep-th/0212279].
  
   \bibitem{ahs}
 H.~Awata, S.~Hirano and M.~Shigemori, ``The Partition Function of ABJ Theory,''
  Prog.\  Theor.\  Exp.\  Phys.\ , 053B04 (2013)
  [arXiv:1212.2966].
  
  \bibitem{honda}
  M.~Honda, ``Direct derivation of "mirror" ABJ partition function,''
  JHEP {\bf 1312}, 046 (2013)
  [arXiv:1310.3126 [hep-th]].

      \bibitem{honda-o}
 M.~Honda and K.~Okuyama, ``Exact results on ABJ theory and the refined topological string,''
  JHEP {\bf 1408}, 148 (2014)
  [arXiv:1405.3653 [hep-th]].
 
\bibitem{dmp}
N.~Drukker, M.~Mari\~no and P.~Putrov, ``From weak to strong coupling in ABJM theory,''
  Commun.\ Math.\ Phys.\  {\bf 306}, 511 (2011)
  [arXiv:1007.3837 [hep-th]].

\bibitem{akmv-cs}
   M.~Aganagic, A.~Klemm, M.~Mari\~no and C.~Vafa, ``Matrix model as a mirror of Chern-Simons theory,''
  JHEP {\bf 0402}, 010 (2004)
  [hep-th/0211098].
  
  \bibitem{gv}
R.~Gopakumar and C.~Vafa, ``On the gauge theory / geometry correspondence,''
  Adv.\ Theor.\ Math.\ Phys.\  {\bf 3}, 1415 (1999)
  [hep-th/9811131].
  
 \bibitem{mmcs}
 M.~Mari\~no, ``Chern-Simons theory, matrix integrals, and perturbative three manifold invariants,''
  Commun.\ Math.\ Phys.\  {\bf 253}, 25 (2004)
  [hep-th/0207096].

\bibitem{hy}
 N.~Halmagyi and V.~Yasnov, ``The Spectral curve of the lens space matrix model,''
  JHEP {\bf 0911}, 104 (2009)
  [hep-th/0311117].

\bibitem{hoy}
N.~Halmagyi, T.~Okuda and V.~Yasnov, ``Large N duality, lens spaces and the Chern-Simons matrix model,''
  JHEP {\bf 0404}, 014 (2004)
  [hep-th/0312145].



  \bibitem{hkp}
 M.~X.~Huang, A.~Klemm and M.~Poretschkin, ``Refined stable pair invariants for E-, M- and $[p, q]$-strings,''
  JHEP {\bf 1311}, 112 (2013)
  [arXiv:1308.0619 [hep-th]].
  
  \bibitem{hkrs}
   M.~x.~Huang, A.~Klemm, J.~Reuter and M.~Schiereck,
  ``Quantum geometry of del Pezzo surfaces in the Nekrasov-Shatashvili limit,''
  JHEP {\bf 1502}, 031 (2015)
  [arXiv:1401.4723 [hep-th]].
 
  \bibitem{kkv}  
   S.~H.~Katz, A.~Klemm and C.~Vafa, ``Geometric engineering of quantum field theories,''
  Nucl.\ Phys.\ B {\bf 497}, 173 (1997)
  [hep-th/9609239].
 
 \bibitem{faddeev}
L.~D. Faddeev, ``Discrete {H}eisenberg-{W}eyl group and modular group,"
  Lett. Math. Phys. \textbf{34},  249 (1995).

\bibitem{fk}
L.~D. Faddeev and R.~M. Kashaev, ``Quantum dilogarithm," Modern
  Phys. Lett. A \textbf{9}, 427 (1994).
  
\bibitem{fkv}
L.~D.~Faddeev, R.~M.~Kashaev and A.~Y.~Volkov, ``Strongly coupled quantum discrete Liouville theory. 1. Algebraic approach and duality,''
  Commun.\ Math.\ Phys.\  {\bf 219}, 199 (2001)
  [hep-th/0006156].

\bibitem{faddeev-penta} 
   L.~D.~Faddeev,
  ``Current - like variables in massive and massless integrable models,''
  In *Varenna 1994, Quantum groups and their applications in physics* 117-135
  [hep-th/9408041].

 \bibitem{zamo}
  A.~B.~Zamolodchikov, ``Painlev\'e III and 2-d polymers,''
  Nucl.\ Phys.\ B {\bf 432}, 427 (1994)
  [hep-th/9409108].
   
   \bibitem{tw}
 C.~A.~Tracy and H.~Widom, ``Proofs of two conjectures related to the thermodynamic Bethe ansatz,''
  Commun.\ Math.\ Phys.\  {\bf 179}, 667 (1996)
  [solv-int/9509003].
  
  \bibitem{simon} 
 B. Simon, {\it Trace ideals and their applications}, second edition, American Mathematical Society, Providence, 2000. 
 
\bibitem{kwy2}
A.~Kapustin, B.~Willett and I.~Yaakov, ``Nonperturbative Tests of Three-Dimensional Dualities,''
  JHEP {\bf 1010}, 013 (2010)
  [arXiv:1003.5694 [hep-th]].



 \bibitem{kostovon}
  I.~K.~Kostov, ``O($n$) Vector Model on a Planar Random Lattice: Spectrum of Anomalous Dimensions,''
  Mod.\ Phys.\ Lett.\ A {\bf 4}, 217 (1989).
  
  \bibitem{ks}
  I.~K.~Kostov and M.~Staudacher, ``Multicritical phases of the $O(n)$ model on a random lattice,''
  Nucl.\ Phys.\ B {\bf 384}, 459 (1992)
  [hep-th/9203030].
  
  \bibitem{AS}
 M. Abramowitz and I. Stegun, {\it Handbook of Mathematical Functions with Formulas, Graphs, and Mathematical Tables}, New York, Dover.   

 \bibitem{kt}
I.~R.~Klebanov and A.~A.~Tseytlin, ``Entropy of near extremal black p-branes,''
  Nucl.\ Phys.\ B {\bf 475}, 164 (1996)
  [hep-th/9604089].

 \bibitem{hkpt}
 C.~P.~Herzog, I.~R.~Klebanov, S.~S.~Pufu and T.~Tesileanu, ``Multi-Matrix Models and Tri-Sasaki Einstein Spaces,''
  Phys.\ Rev.\ D {\bf 83}, 046001 (2011)
  [arXiv:1011.5487 [hep-th]].

 \bibitem{kwy}
A.~Kapustin, B.~Willett and I.~Yaakov, ``Exact Results for Wilson Loops in Superconformal Chern-Simons Theories with Matter,''
  JHEP {\bf 1003}, 089 (2010)
  [arXiv:0909.4559 [hep-th]].
  
 
    \bibitem{abjm}
 O.~Aharony, O.~Bergman, D.~L.~Jafferis and J.~Maldacena, ``N=6 superconformal Chern-Simons-matter theories, M2-branes and their gravity duals,''
  JHEP {\bf 0810}, 091 (2008)
  [arXiv:0806.1218 [hep-th]].
    
 \bibitem{abj}
  O.~Aharony, O.~Bergman and D.~L.~Jafferis,
  ``Fractional M2-branes,''
  JHEP {\bf 0811} (2008) 043
  [arXiv:0807.4924 [hep-th]].
  %
  
   \bibitem{abj-moriyama}
  S.~Matsumoto and S.~Moriyama, ``ABJ Fractional Brane from ABJM Wilson Loop,''
  JHEP {\bf 1403}, 079 (2014)
  [arXiv:1310.8051 [hep-th]].

  \bibitem{mp-abjm}
  M.~Mari\~no and P.~Putrov, ``Exact Results in ABJM Theory from Topological Strings,''
  JHEP {\bf 1006}, 011 (2010)
  [arXiv:0912.3074 [hep-th]].
  
  \bibitem{yhu}
Y. Hatsuda, unpublished. 

      \bibitem{hanada}
 M.~Hanada, M.~Honda, Y.~Honma, J.~Nishimura, S.~Shiba and Y.~Yoshida, 
 ``Numerical studies of the ABJM theory for arbitrary N at arbitrary coupling constant,''
  JHEP {\bf 1205}, 121 (2012)
  [arXiv:1202.5300 [hep-th]].

\bibitem{ho}
Y.~Hatsuda and K.~Okuyama, ``Probing non-perturbative effects in M-theory,''
  JHEP {\bf 1410}, 158 (2014)
  [arXiv:1407.3786 [hep-th]].
  
\bibitem{witten}
E.~Witten, ``Quantum Field Theory and the Jones Polynomial,''
  Commun.\ Math.\ Phys.\  {\bf 121}, 351 (1989).

 \bibitem{kota}
 Y.~Hatsuda and K.~Okuyama, ``Resummations and Non-Perturbative Corrections,''
  JHEP {\bf 1509}, 051 (2015)
  [arXiv:1505.07460 [hep-th]].
  
\bibitem{ruben}
 R.~L.~Mkrtchyan, ``Nonperturbative universal Chern-Simons theory,''
  JHEP {\bf 1309}, 054 (2013)
  [arXiv:1302.1507 [hep-th]].


\bibitem{gv2}
R.~Gopakumar and C.~Vafa, ``M theory and topological strings. 2.,''
  [hep-th/9812127].
  
     \bibitem{abk}
  M.~Aganagic, V.~Bouchard and A.~Klemm, ``Topological Strings and (Almost) Modular Forms,''
  Commun.\ Math.\ Phys.\  {\bf 277}, 771 (2008)
  [hep-th/0607100].

\bibitem{bt}
A.~Brini and A.~Tanzini, ``Exact results for topological strings on resolved Y**p,q singularities,''
  Commun.\ Math.\ Phys.\  {\bf 289}, 205 (2009)
  [arXiv:0804.2598 [hep-th]].

\bibitem{hkr}
B.~Haghighat, A.~Klemm and M.~Rauch, ``Integrability of the holomorphic anomaly equations,''
  JHEP {\bf 0810}, 097 (2008)
  [arXiv:0809.1674 [hep-th]].
  
\bibitem{gl}
B.~R.~Greene and C.~I.~Lazaroiu, ``Collapsing D-branes in Calabi-Yau moduli space. 1.,''
  Nucl.\ Phys.\ B {\bf 604}, 181 (2001)
  [hep-th/0001025].
  
 \bibitem{dlo}
X.~De la Ossa, B.~Florea and H.~Skarke, ``D-branes on noncompact Calabi-Yau manifolds: K theory and monodromy,''
  Nucl.\ Phys.\ B {\bf 644}, 170 (2002)
  [hep-th/0104254].
  
  \bibitem{gvcone}
D.~Ghoshal and C.~Vafa, ``$c = 1$ string as the topological theory of the conifold,''
  Nucl.\ Phys.\ B {\bf 453}, 121 (1995)
  [hep-th/9506122].
  

\bibitem{hk}
 M.~x.~Huang and A.~Klemm, ``Holomorphic Anomaly in Gauge Theories and Matrix Models,''
  JHEP {\bf 0709}, 054 (2007)
  [hep-th/0605195].
  
   \bibitem{mmnp}
  M.~Mari\~no, ``Nonperturbative effects and nonperturbative definitions in matrix models and topological strings,''
  JHEP {\bf 0812}, 114 (2008)
  [arXiv:0805.3033 [hep-th]].
  
  \bibitem{rrr}
   R.~Couso-Santamaria, R.~Schiappa and R.~Vaz, ``Finite N from Resurgent Large N,''
  Annals Phys.\  {\bf 356}, 1 (2015)
  [arXiv:1501.01007 [hep-th]].
  
  \bibitem{cesv1} R.~Couso-Santamaria, J.~D.~Edelstein, R.~Schiappa and M.~Vonk, ``Resurgent Transseries and the Holomorphic Anomaly,''
  Annales Henri Poincare {\bf 17}, no. 2, 331 (2016) [arXiv:1308.1695 [hep-th]].
  
  \bibitem{cesv2} 
  R.~Couso-Santamaria, J.~D.~Edelstein, R.~Schiappa and M.~Vonk, 
  ``Resurgent Transseries and the Holomorphic Anomaly: Nonperturbative Closed Strings in Local $\IC\IP^2$,'' Commun.\ Math.\ Phys.\  {\bf 338}, no. 1, 285 (2015)
  [arXiv:1407.4821 [hep-th]].

  \bibitem{gp}
  M.~Gaudin and V.~Pasquier, ``The periodic Toda chain and a matrix generalization of the bessel function's recursion relations,''
  J.\ Phys.\ A {\bf 25}, 5243 (1992).
    
      \bibitem{nekrasov}
   N.~A.~Nekrasov, ``Seiberg-Witten prepotential from instanton counting,''
  Adv.\ Theor.\ Math.\ Phys.\  {\bf 7}, 831 (2004)
  [hep-th/0206161].
  
  \bibitem{krefl}
   D.~Krefl, ``Non-Perturbative Quantum Geometry II,''
  JHEP {\bf 1412}, 118 (2014)
  [arXiv:1410.7116 [hep-th]].
    
     \bibitem{dunne}
  G.~Basar and G.~V.~Dunne, ``Resurgence and the Nekrasov-Shatashvili limit: connecting weak and strong coupling in the Mathieu and Lam\'e systems,''
  JHEP {\bf 1502}, 160 (2015)
  [arXiv:1501.05671 [hep-th]].
  
     \bibitem{kpt} 
 A.~K.~Kashani-Poor and J.~Troost, ``Pure $ \mathcal{N}=2 $ super Yang-Mills and exact WKB,''
  JHEP {\bf 1508}, 160 (2015) [arXiv:1504.08324 [hep-th]].

 \bibitem{cmsy}
   C.~M.~Chang, S.~Minwalla, T.~Sharma and X.~Yin, ``ABJ Triality: from Higher Spin Fields to Strings,''
  J.\ Phys.\ A {\bf 46}, 214009 (2013)
  [arXiv:1207.4485 [hep-th]].
   
   
  \bibitem{hhos}
  S.~Hirano, M.~Honda, K.~Okuyama and M.~Shigemori, ``ABJ Theory in the Higher Spin Limit,''
  arXiv:1504.00365 [hep-th].

\bibitem{ak}
J.~Ellegaard Andersen and R.~Kashaev, ``A TQFT from Quantum Teichm\"uller Theory,''
  Commun.\ Math.\ Phys.\  {\bf 330}, 887 (2014)
  [arXiv:1109.6295 [math.QA]].
  
  \bibitem{mcintosh}
R. J. Mcintosh, ``Some asymptotic formulae for q-shifted factorials," The Ramanujan Journal {\bf 3} (1999) 205.
  
  \bibitem{kasteleyn}
 P. W. Kasteleyn, ``The statistics of dimers on a lattice: I. The number of dimer arrangements on a quadratic lattice," Physica {\bf 27} (1961) 1209.

\bibitem{david}   
  D. Cimasoni, ``The geometry of dimer models," arXiv:1409.4631.
  
 \end{thebibliography}
\end{document}